\documentclass [11pt]{article}
\usepackage{latexsym}

\newcommand{\beq}{\begin{equation}}
\newcommand{\eeq}{\end{equation}}
\newcommand{\bea}{\begin{eqnarray}}
\newcommand{\eea}{\end{eqnarray}}
\newcommand{\barr}[1]{\begin{array}}
\newcommand{\earr}{\end{array}}
\newtheorem{theorem}{Theorem}[section]
\newtheorem{definition}{Definition}[section]
\newtheorem{proposition}{Proposition}[section]
\newcommand{\bdf}{\begin{definition}}
\newcommand{\edf}{\end{definition}}

\newcounter{abc}
\newenvironment{zoznamrom}{\setcounter{abc}{0}\begin{list}{\roman{abc})}
                       {\usecounter{abc}}}{\end{list}}

\newcounter{zozal}
\newenvironment{zoznamAlph}{\setcounter{zozal}{0}\begin{list}{\Alph{zozal})}
                       {\usecounter{zozal}}}{\end{list}}

\newcounter{znum}
\newenvironment{zoznamnum}{\setcounter{znum}{0}\begin{list}{\arabic{znum}.}
                       {\usecounter{znum}}}{\end{list}}

\def\kp{k_{+}} 
\def\km{k_{-}} 
\def\kz{k_{0}}
\def\mR{\mathbf R}
\def\pq{(p,q)}

\begin{document}

\begin{titlepage}
\topmargin=2.0in
\begin{center}
{\LARGE Maximal Abelian Subgroups of the Isometry and \\ [0.2cm]
Conformal Groups of Euclidean and Minkowski Spaces}

\vskip 2.5em

{\large Z.~Thomova and P.~Winternitz \\
        Centre de recherches math{\' e}matiques \\
        Universit{\'e} de Montr{\' e}al \\
        Case postale 6128, succursale centre-ville \\
        Montr{\'e}al (Qu{\'e}bec) H3C 3J7, Canada}
\end{center}
\end{titlepage}        


\begin{abstract}
The maximal Abelian subalgebras of the Euclidean $e(p,0)$ and pseudoeuclidean $e(p,1)$ 
Lie algebras are classified into conjugacy classes under the action of the corresponding 
Lie groups $E(p,0)$ and $E(p,1)$, and also under the conformal groups $O(p+1,1)$ and 
$O(p+1,2)$, respectively. The results are presented in terms of decomposition theorems. 
For $e(p,0)$ orthogonally indecomposable MASAs exist only for $p=1$ and $p=2$. For 
$e(p,1)$, on the other hand, orthogonally indecomposable MASAs exist for all values of 
$p$. The results are used to construct new coordinate systems in which wave equations 
and Hamilton-Jacobi equations allow the separation of variables.

\end{abstract}
\newpage

\bibliographystyle{unsrt}


\section {Introduction}

The stage of much of mathematical physics is the real flat space $\mR^n$ with a 
nondegenerate indefinite metric of signature $\pq$. We shall denote this space $M\pq$ 
with $p+q=n$. The isometry group of this space is the pseudoeuclidean group $E\pq$ and 
the conformal group is $C\pq \sim O(p+1,q+1)$ (the pseudoorthogonal group in $p+q+2$ 
dimensions, acting locally and nonlinearly on $M\pq$).

The purpose of this article is to present a classification of the maximal 
Abelian subalgebras (MASAs) of the real Euclidean and pseudoeuclidean Lie 
algebras $e(p,0) \equiv e(p)$ and $e(p,1)$. The classification is first 
performed with respect to conjugation under the corresponding Lie groups 
$E(p,0) \equiv E(p)$ and $E(p,1)$, respectively, and it also provides a classification of 
the connected maximal Abelian subgroups of the corresponding groups $E(p)$ and $E(p,1)$. 
We also present a classification of MASAs of the corresponding conformal algebras 
$c(p,0) \sim o(p+1,1)$ and $c(p,1) \sim o(p+1,2)$ under the corresponding groups 
$O(p+1,1)$ and $O(p+1,2)$. This classification is used to show (for $q=0$ or $1$) 
which MASAs of $e\pq$ are also MASAs of $o(p+1,q+1)$ and which MASAs that are 
inequivalent under $E\pq$ are nevertheless mutually conjugated under the larger 
conformal group $O(p+1,q+1)$.

The classification of MASAs of $e\pq$ ($q=0,1$) will be used to address a 
physical problem: the separation of variables in Laplace-Beltrami and Hamilton-Jacobi 
equations in the corresponding spaces $M\pq$.

The motivation for our study of subgroups of Lie groups and subalgebras of 
Lie algebras is multifold. For instance, consider any physical problem leading 
to a system of differential, difference, algebraic, integral or other equations. 
Let the set of all solutions of the system be invariant under some Lie group $G$, 
the "symmetry group". Special solutions, corresponding to special boundary, 
or initial conditions, can be constructed as "invariant solutions", invariant under 
some subgroup of the group $G$ ~\cite{olver,kluwer}. For linear equations, or 
for Hamilton-Jacobi type equations, solutions obtained by separation of variables 
are examples of invariant solutions. While all types of subgroups $G_{0} \subset G$ 
are relevant to this problem, Abelian subgroups provide particularly simple reductions 
and particularly simple coordinate systems. Indeed, each one-dimensional subalgebra 
of an Abelian symmetry algebra will provide an "ignorable" variable
~\cite{frisw,wsmor,miller,kalnins,milpatw,eisen}, 
i.e. a variable that does not figure in the metric 
tensor (a "cyclic" variable in classical mechanics).

Another example of the application of maximal 
Abelian subgroups of an invariance 
group is in any quantum theory, where Abelian subalgebras provide sets of 
commuting operators that characterize states of a physical system. The system 
itself is characterized by the Casimir operators of the group $G$. Complete 
information about possible quantum numbers would be provided by constructing 
MASAs of the enveloping algebra of the Lie algebra $L$ of $G$. MASAs of the Lie
 algebra itself provide additive quantum numbers.

A third application is in the theory of integrable systems, both finite and 
infinite dimensional, where MASAs of any underlying Lie algebra provide 
integrals of motion in involution, commuting flows, and other basic information 
about the systems.

A series of earlier articles was devoted to MASAs of the classical Lie algebras, 
such as $sp(2n,R)$ and $sp(2n,C)$ ~\cite{sp_C}, $su(p,q)$ ~\cite{su_pq}, 
$so(n,C)$ ~\cite{verc} and $so(p,q)$ ~\cite{verop}. Special roles amongst all MASAs 
of simple and semisimple Lie algebras are played by Cartan subalgebras on one hand 
and maximal Abelian nilpotent algebras (MANSs), on the other. The Cartan 
subalgebras are their own normalizers \cite{jacobson} and consist entirely of 
nonnilpotent elements. For a complex semisimple Lie algebra there is, up to 
conjugacy, only one Cartan subalgebra. For real semisimple Lie algebras they 
were classified by 
Kostant \cite{kostant}, and Sugiura \cite{sugiura}. Maximal Abelian nilpotent 
subalgebras consist entirely of nilpotent elements (represented by nilpotent 
matrices in any finite dimensional representation). They were studied by Kravchuk 
for $sl(n,C)$ and his results are summed up in book form \cite{suprtysh}. Maltsev 
obtained all MANSs of maximal dimension for the simple Lie algebras \cite{maltsev}. 
Those of minimal dimension have also been studied \cite{laffey}.

More recently, the study of MASAs was extended to the inhomogeneous classical 
Lie algebras, or finite dimensional affine Lie algebras, starting from the 
complex Euclidean Lie algebras $e(n,C)$~\cite{kal}. 

The next natural step is 
to consider the real Euclidean and pseudoeuclidean algebras $e(p,q)$ for 
$p \geq q \geq 0$.
This study is initiated in the present article 
where we concentrate on the values $q=0$ and $1.$ On one hand, these 
are the most important ones in physical applications, since they include the 
Lie algebras of the groups of motions $E(p)$ of 
Euclidean spaces and $E(p,1)$ of Minkowski spaces. 
On the other, they are the simplest ones to treat, so all results are 
entirely explicit. The general case of $q\geq 2$ will be treated separately and 
is  more complicated from the mathematical point of view. 

The classification strategy and some general results on the MASAs of $e(p,q)$ 
are presented in Section 2. The real Euclidean algebra $e(p)$ is treated 
in Section 3 where we also list the MASAs of $o(p,1)$ and
classification of MASAs of $e(p)$ under
 the action of the group $O(p+1,1)$. Section 4 then 
treats MASAs of $e(p,1)$.
Section 5 lists results on MASAs of $o(p,2)$ and the classification of MASAs 
of $e(p,1)$ 
under the action of the conformal group $O(p+1,2)$ of compactified 
Minkowski space $M(p,1)$. In other words, certain MASAs not conjugated 
under $E(p,1)$, are conjugated under the larger group $O(p+1,2)$. 
MASAs of $e(p,1)$ are used in Section 6 to obtain maximal Abelian 
subgroups of $E(p,1)$. These in turn provide us with all separable 
coordinate systems in Minkowski space $M(p,1)$ with a maximal number 
of ignorable variables. Some conclusions are drawn in Section 7.



\section {General formulation}

\subsection {Some definitions}

We will be classifying maximal Abelian subalgebras of the pseudoeuclidean Lie 
algebra $e(p,q)$ into conjugacy classes under the action of the pseudoeuclidean 
Lie group $E(p,q)$. A convenient realization of this algebra and this group is 
by real matrices $Y$ and $H$ , satisfying
\bea 
Y(X,\alpha) \equiv Y =\left( \begin{array}{cc} X  & \alpha \\
                                               0 & 0 
                              \end{array} \right),
\qquad X \in \mathbf{R}^{n \times n}, \qquad \alpha \in \mathbf{R}^{n \times 1} \label{H}
\eea
\bea
H=\left( \begin{array}{cc} G & a \\
                            0 & 1 
           \end{array} \right),                                   
\qquad G \in \mathbf{R}^{n \times n},\qquad a \in \mathbf{R}^{n \times 1}, \label{G}
\eea
respectively, where $X$ and $G$ satisfy
$$
XK+KX^{T}=0, \quad GKG^{T}=K,
$$
\beq
K=K^{T} \in \mathbf{R}^{n \times n}, \quad n=p+q, \quad det K \not= 0, \label{2.3}
\eeq
$$
sgn K=(p,q), \quad p \geq q \geq 0,
$$
respectively. Here $sgn K$ denotes the signature of $K$, with $p$ the number 
of positive eigenvalues of $K$ and $q$ the number of negative ones. We shall 
also make use of an "extended" matrix $K_{e} \in \mathbf{R}^{(n+1) \times (n+1)}$ 
satisfying
\bea 
K_{e}= \left( \begin{array}{cc} K & 0 \\
                                 0 & 0_{1} 
                \end{array} \right), \quad
YK_{e}+K_{e}Y^{T}=0. \label{Ke}
\eea

A convenient basis for the algebra $e(p,q)$ is provided by $n$ translations
$P_{\mu}$ and $n(n-1)/2$ rotations and pseudorotations $L_{\mu \nu}$. The commutation 
relations for 
this basis are
\bea
{[L_{ik},L_{ab}]} & = & \delta_{ka}L_{ib}-\delta_{kb}L_{ia}-
\delta_{ia}L_{kb}+\delta_{ib}L_{ka} \nonumber \\
{[L_{\alpha\beta},L_{\gamma\delta}]}&=&\delta_{\beta\gamma}L_{\alpha\delta}
-\delta_{\beta\delta}L_{\alpha\gamma}-\delta_{\alpha\gamma}L_{\beta\delta}
+\delta_{\alpha\delta}L_{\beta\gamma} \nonumber \\
{[L_{ik},L_{a\beta}]} & = & \delta_{ka}L_{i\beta}-
\delta_{ia}L_{k\beta}  \label{comm} \\
{[L_{i \alpha},L_{\beta\gamma}]}&=& \delta_{\alpha\beta}L_{i \gamma}
-\delta_{\alpha\gamma}L_{i \beta} \nonumber \\
{[L_{a\beta},L_{i\mu}]} & = & \delta_{\beta\mu}L_{ai}+
\delta_{ai}L_{\beta\mu} \nonumber 
\eea
where $i,k,a,b \leq p$ and $p < \alpha, \beta, \gamma, \delta, \mu \leq q$,
\bea
{[P_{\alpha},L_{\mu\nu}]} & = & g_{\alpha\mu}P_{\nu}-g_{\alpha\nu}P_{\mu} \label{comm1} \\
{[P_{\mu},P_{\nu}]} & = & 0 \nonumber
\eea
for $0 < \alpha, \mu, \nu \leq p+q$,
\begin{eqnarray*}
g_{11}=g_{22}= \ldots =g_{pp}=-g_{p+1,p+1}= \ldots =-g_{p+q,p+q}=1 \\
g_{\mu \nu}=0 \quad {\rm for} \quad \mu \not= \nu.
\end{eqnarray*}

A standard realization of this basis in terms of differential operators is given by
\bea
P_{\mu}= {\partial \over \partial x_{\mu}}, \quad L_{ik}=x_{i}{\partial \over \partial 
x_{k}}-x_{k}{\partial \over \partial x_{i}} \label{diffop}
\eea
for $1 \leq i <k \leq p$ or $p+1 \leq i <k \leq p+q$ and
\begin{eqnarray*}
 L_{ik}=-(x_{k}{\partial \over \partial x_{i}}+x_{i}{\partial \over \partial x_{k}}), 
\quad 1 \leq i \leq p, \quad p+1\leq k\leq p+q.
\end{eqnarray*}

From the above discussion we see that the pseudoeuclidean Lie algebra is the semidirect sum of the pseudoorthogonal Lie algebra $o(p,q)$ and an Abelian algebra $T(n)$ of translations.

Since $T(n)$ is an ideal in $e(p,q)$, we can consider the adjoint representation of $o(p,q)$ on $T(n)$. Abusing notation, we use the same letters ${P_{1}, \ldots,P_{p},P_{p+1},\ldots, P_{p+q}}$ for basis vectors in this representation. The metric tensor $g_{\mu\nu}$ defined above provides an invariant scalar product on the representation space
\beq
 (P,Q)=g_{\mu\nu}P_{\mu}Q_{\nu}.
\eeq
We shall call vectors satisfying $P^{2}>0$, $P^{2}<0$ and $P^{2}=0 \, (P\not=0)$ positive length, negative length and isotropic, respectively.

We also need to define some basic algebraic concepts.
\bdf
The centralizer $cent(L_{0},L)$ of a Lie algebra $L_{0}\in L$ is a subalgebra of L 
consisting of all elements in L, commuting elementwise with $L_{0}$
\beq
cent(L_{0},L)=\{e \in L|[e,L_{0}]=0\}.
\eeq
\edf

\bdf
A maximal Abelian subalgebra $L_{0}$ (MASA) of L is an Abelian subalgebra, equal
to its centralizer
\beq
[L_{0},L_{0}]=0, \, \, \, \, \,  cent(L{_0},L)=L_{0}.
\eeq
\edf

\bdf
A splitting subalgebra $L_{0}$ of the semidirect sum
\beq
L=F \triangleright N,\, \, [F,F]\subseteq F, \, \, [F,N]\subseteq N, \, \,  
[N,N]\subseteq N
\eeq
is itself a semidirect sum of a subalgebra of F and a subalgebra of N
\beq
L_{0}=F_{0}\triangleright N_{0},\, \, \, F_{0}\subseteq F, \, \, \, N_{0}\subseteq N 
\eeq
(or conjugate to such a semidirect sum).
\edf

All other subalgebras of $L=F \triangleright N$ are called 
{\it nonsplitting subalgebras}.

An {\it Abelian splitting subalgebra} of $L=F\triangleright N$ is a direct sum
\beq
L_{0}=F_{0}\oplus N_{0},\, \, \, \, F_{0}\subseteq F, \, \, \, \,
N_{0}\subseteq N.
\eeq

\bdf
A maximal Abelian nilpotent subalgebra (MANS) M of a Lie algebra L is a MASA, 
consisting entirely of nilpotent elements, i.e. it satisfies
\beq
[M,M]=0, \, \, \, \, \, \left[ \left[ [L,M]M\right] \ldots \right]_{m}=0
\eeq
for some finite number m (we commute M with L $m$-times).
\edf

Let us now consider the pseudoeuclidean space $M(p,q)$, i.e. ${\mathbf R}^n, 
n=p+q$ with an invariant quadratic form given by the matrix K of eq.(\ref{2.3})
\bea
ds^2=dx^TKdx. \label{2.14}
\eea
The group and Lie algebra actions are given by
\bea
x'=Gx+a, \qquad \qquad x'=Xx+\alpha \label{2.15}
\eea
respectively, with $(X,\alpha)$ and $(G,a)$ as in eq.(\ref{H}) and (\ref{G}).

\begin{definition}
A subalgebra $L_0 \subset e(p,q)$ is orthogonally decomposable if it preserves
an orthogonal decomposition of $M(p,q)$ 
\bea
M(p,q) = M(p_1,q_1) \oplus M(p_2,q_2), \qquad p_1+p_2=p, \qquad q_1+q_2=q
\label{2.16a}
\eea
into two (or more) nonempty subspaces. It is called orthogonally indecomposable otherwise.
\end{definition}

\subsection {Classification strategy}

The classification of MASA's of $e(p,q)$ is based on the fact that $e(p,q)$ is 
the semidirect sum of the Lie algebra $o(p,q)$ and an 
Abelian ideal $T(n)$ 
(the translations). We use here a modification of a  procedure described earlier ~\cite{kal} 
for $e(n, C)$. We proceed in five steps.

\begin{zoznamnum}
\item Classify subalgebras $T(k_{+},k_{-},k_{0})$ of $T(n)$. They are 
characterized by a triplet of nonnegative integers $(k_{+},k_{-},k_{0})$ where $k_{+},k_{-}$ and $k_{0}$ 
are the numbers of positive, negative and isotropic vectors in an orthogonal basis, 
respectively.
\item Find the centralizer $C(k_{+},k_{-},k_{0})$ of $T(k_{+},k_{-},k_{0})$ in 
$o(p,q)$
\bea C(k_{+},k_{-},k_{0})=\{X \in o(p,q)|[X,T(k_{+},k_{-},k_{0})]=0\}. \eea
\item Construct all MASAs of $C(k_{+},k_{-},k_{0})$ and classify them under the 
action of normalizer $Nor[T(k_{+},k_{-},k_{0}),G]$ of $T(k_{+},k_{-},k_{0})$
in the group $G \sim E(p,q)$.
\item Obtain a list of  splitting MASAs of $e(p,q)$ by forming the direct sums
\bea C(k_{+},k_{-},k_{0}) \oplus T(k_{+},k_{-},k_{0}) \eea
and dropping all such algebras that are not maximal from the list.
\item Complement the basis of $T(k_{+},k_{-},k_{0})$ to a basis of $T(n)$ in each 
case and construct all nonsplitting MASAs. The procedure is described below in Section 4.2.
\end{zoznamnum}

This general strategy can also be expressed in terms of sets of matrices of the form
(\ref{H}), \ldots ,(\ref{Ke}).

The subalgebra $T(k_{+},k_{-},k_{0})$ can be represented by the matrices
\bea
\Pi=\left( \begin{array}{cccccl} 0_{k_{0}} & \, & \, & \, & \, & \xi \\
                                  \, & 0_{p+q-2k_{0}-k_{+}-k_{-}} & \, & \, & \, & 0 \\
                                  \, & \, &  0_{k_{0}} & \, & \, & 0 \\
                                   \, & \, & \, &  0_{k_{+}} & \, & x \\
                                   \, & \, & \, & \, &  0_{k_{-}} & y \\
                                    \, & \, & \, & \, & \, & 0_{1}
            \end{array} \right),  \label{2.16}
\eea
\bea
K_{e}=\left( \begin{array}{cccccl} \, & \, & I_{k_{0}} & \, & \, & 0 \\
                                  \, & K_{0} & \, & \, & \, & \vdots \\
                                   I_{k_{0}} & \, &  \, & \, & \, & \vdots \\
                                   \, & \, & \, &  I_{k_{+}} & \, & 0 \\
                                   \, & \, & \, & \, &  -I_{k_{-}} & 0 \\
                                    \, & \, & \, & \, & \, & 0_{1}
            \end{array} \right),  \label{2.17} 
\eea
where $K_{0}$ has  signature $(p\!-\!k_{+}\!-\!k_{0},q\!-\!k_{-}\!-\!k_{0})$.

The centralizer $C(k_{+},k_{-},k_{0})$ of $T(k_{+},k_{-},k_{0})$ will 
then be represented by the block diagonal matrices
\bea
C=\left( \begin{array}{cccc} \tilde{M} & \, & \, & \, \\
                                     \, & 0_{k{+}} & \, & \, \\
                                    \, & \, & 0_{k{-}} & \, \\
                                     \, & \, & \, & 0_{1}
         \end{array} \right), \qquad
\tilde{M}= \left( \begin{array}{ccc} 0_{k_{0}} & \tilde A & \tilde Y \\
                                         0 & \tilde S & - \tilde K \tilde A^{T} \\
                                         0 & 0 & 0_{k_{0}}
                 \end{array} \right)  \label{2.18} \\ 
\tilde Y = -\tilde Y^{T}, \qquad  \tilde S\tilde K+\tilde K \tilde S^{T}=0. \nonumber 
\eea

The Lie algebra of matrices $\{\tilde{M}\}$ represents a subalgebra  of 
$o(p\!-\!k_{+},q\!-\!k_{-})$ and we need to classify the MASAs of 
$o(p\!-\!k_{+},q\!-\!k_{-})$ contained in $\{\tilde{M}\}$.
Such  MASAs were studied elsewhere \cite{verop} and we shall recall some 
basic facts here.

A MASA of $o(p,q)$ is characterized by a set of matrices $X$ and a "metric" 
matrix $K$, satisfying eq.~(\ref{2.3}). A MASA can be orthogonally indecomposable 
(OID), or orthogonally decomposable (OD). If it is OD, we decompose it, 
{\it i.e.} transform it, together with $K$, into block diagonal form. Each block 
is an OID MASA of some $o(p_{i},q_{i})$, $\sum p_{i}=p, \sum q_{i}=q.$  At 
most one of the blocks is a MANS.

From the above we can see that the MASA of $e(p,q)$ will have the following general form
\bea
M= \left( \begin{array}{ccccccc} 0_{k_{0}} & A & Y & \, & \, & \, & \xi \\
                                    \, & S & -K_{p_1 q_1}A^{T} & \, & \, & \, & \, \\
                                     \, & \, &  0_{k_{0}} & \, & \, & \, & \, \\
                                     \, & \, & \, & M_{1} & \, & \, & \, \\
                                     \, & \, & \, & \, & 0_{k_{+}} & \, & x \\
                                      \, &  \, & \, & \, & \, & 0_{k_{-}} & y \\
                                      \, & \, &  \, & \, & \, & \, & 0_{1} 
                    \end{array} \right), 
\eea
\bea
K_{e}=\left( \begin{array}{ccccccc} \, & \, & I_{k_{0}} & \, & \, & \, & \, \\
                                    \, & K_{p_1 q_1} & \, & \, & \, & \, & \, \\
                                     I_{k_{0}} & \, & \, & \, & \, & \, & \, \\
                                     \, & \, & \, & K_{p_2 q_2} & \, & \, & \, \\
                                     \, & \, & \, & \, & I_{k_{+}} & \, & \, \\
                                      \, &  \, & \, & \, & \, & -I_{k_{-}} & \, \\
                                      \, & \, &  \, & \, & \, & \, & 0_{1} 
                    \end{array} \right) 
\eea
where $M_{1}$ is a MASA of $o(p_2,q_2)$ not containing a MANS, 
$p=p_1+p_2+k_{+}+k_{0}$ and $q=q_1+q_2+k_{-}+k_{0}$. The MASA $M_{1}$ can be 
absent (when $p_2=q_2=0$). It may be orthogonally decomposable.

The block
\bea
M_0=\left( \begin{array}{ccc} 0_{k_0} & A & Y \\
                               0 & S & -K_{p_1 q_1}A^T \\
                                0 & 0 & 0_{k_{0}}
            \end{array} \right), \\
            Y+Y^T=0, \qquad S K_{p_1 q_1}+K_{p_1 q_1}S^T=0 \nonumber
\eea
represents a MANS of $o(p_1+k_0,q_1+k_0)$, so $S \in \mR^{(p_1+q_1) 
\times(p_1+q_1)}$ is a 
nilpotent matrix. For $k_0=0$ the MANS $M_0$ is absent.

\subsection{Embedding into the conformal Lie algebra}

The algebra $o(p+1,q+1)$ contains the rotations and pseudorotations $L_{\alpha \beta}$, translations
$P_\mu$, the dilation $D$ and the proper conformal transformations $C_\mu$.
The realization of the additional basis elements in terms differential operators is given by:
\bea \label{conf}
D=x_{\alpha}{\partial \over \partial x_{\alpha}}, \qquad
C_a=g_{aa}x_ax_\alpha{\partial \over \partial x_\alpha}-{1 \over 2}
(x_\alpha g_{\alpha\beta}x_\beta) {\partial \over \partial x_0}.
\eea

They satisfy the following commutation relations:
\bea \label{2.24}
{[P_\mu,C_\alpha]} & = & 2g_{\mu\alpha}D-2g_{\alpha\alpha}L_{\mu\alpha} \nonumber \\
{[C_\alpha,L_{\mu\nu}]} & = &g_{\alpha\mu}C_\mu-g_{\alpha\nu}C_\mu \nonumber \\
{[D,L_{\mu\nu}]} & = & 0 \\
{[P_\mu,D]} & = & P_\mu \nonumber  \\
{[C_\mu,D]} & = & -C_\mu \nonumber
\eea

A matrix representation of $o(p+1,q+1$) is 
\bea \label{2.25}
M_C=\left( \begin{array}{ccc} d & \alpha & 0 \\
                               \beta^T & X_0 & -K_0\alpha^T \\
                               0 & -\beta K_0 & -d
            \end{array} \right), \qquad 
K_C=\left( \begin{array}{ccc} \, & \, & 1 \\
                              \, & K_0 & \, \\
                              1 & \, & \, 
            \end{array} \right),  \\
X_0K_0+K_0X_0^T=0 \nonumber
\eea
where $\alpha, \beta, d, X_0$ represent translations, conformal transformations, 
the dilation, rotations and pseudorotations, respectively. $K_0$ has signature $\pq$. We
have
\beq \label{2.28}
M_CK_C+K_CM_C^T=0.
\eeq

We see that in eq. (\ref{2.25}) the algebra $e\pq$ is embedded as a subalgebra 
of one of the maximal subalgebras of $o(p+1,q+1)$, namely the similitude algebra 
$sim\pq$
obtained by setting $\beta=0$ in (\ref{2.25}). The MASAs of $e\pq$ are thus embedded 
into $o(p+1,q+1)$. In each case we shall determine whether a MASA of $e\pq$ is also  
maximal in $o(p+1,q+1)$. Conversely this representation can be used to determine 
whether a MASA of $o(p+1,q+1)$ is contained in  $e\pq$. Finally, we shall use it 
to establish possible conformal equivalences between MASAs of $e\pq$ that are 
inequivalent under $E\pq$.



\setcounter{equation}{0}
\section{MASA's of e(p,0) and  o(p,1)}
\subsection{Classification of all MASA's of e(p,0) $\equiv$ e(p)}

The metric is  positive definite and, hence, 
a subspace of the translations is completely characterized 
by its dimension. 

A basis for {\textit{e(p)}} is given by $L_{ik}$, $1 \leq i < k \leq p$, and $P_{1}, 
\ldots ,P_{p}$.
\begin{theorem} \label{ep0}
Every MASA of $e(p,0)$ splits into the direct sum $M(k)=F(k) \oplus T(k)$ 
and is 
$E(p,0)$ conjugate to precisely one subalgebra with 
\begin{eqnarray*}
F(k) = \{L_{12}, L_{34}, \ldots, L_{2l-1,2l} \}, \quad 
T(k) = \{ P_{2l+1}, \ldots, P_p \}  
\end{eqnarray*}
where $k$ is such that   $p-k$  is even ($p-k$=2l).
\end{theorem}
{\it Proof.} We take $T(k)=\{P_{p-k+1}, \ldots, P_{p}\}$. Its centralizer
 in $o(p,0)$ is $ o(p-k,0).$ This algebra has just one class of MASAs, 
namely the Cartan subalgebra: \\ 
1. $\tilde F_k=\{L_{12},L_{34} ,\ldots ,L_{p\!-\!k\!-\!1,p\!-\!k}\}$ if $p-k$ 
is even \\
2. $\tilde F_k=\{L_{12},L_{34}, \ldots , L_{p\!-\!k\!-\!2,p\!-\!k\!-\!1}\}$ if $p-k$ 
is odd.\\
The splitting MASA's then would be $T(k) \oplus \tilde F_k$, but for $p-k$  odd, 
the subalgebra is not  maximal. The elements of a nonsplitting MASA would have the
 form $X=L_{a,a+1}+\sum_{j=1}^{p-k} \alpha_{a, j} P_{j}$ where $a=1, 3 \ldots p-k-1$. After
 imposing the commutation relations $[X,Y]=0$ we obtain that all $\alpha_{a, j}=0$. There are no nonsplitting MASA's. \\
\hspace*{\fill} $\Box$


\subsection{MASA's of o(p,1)}
We present here some results from Ref~\cite{verop} on MASA's of  $o(p,1)$.
A MASA of $o(p,1)$ can be
\begin{zoznamnum}
\item Orthogonally decomposable. Two decomposition patterns are possible, namely: 
{\it a)} $l(2,0) \oplus (k,1)$ for $k=0,1, \ldots p-2 \quad (l \geq 1)$ where 
$(k,1)$ is a MANS \\
{\it b)} $(1,1) \oplus (1,0) \oplus l(2,0)$.
\item Orthogonally indecomposable.
Then the MASA  is a MANS of $o(p,1)$.
\end{zoznamnum}

A representative list of $O(p,1)$ conjugacy classes of MANSs of $o(p,1)$ is 
given by the matrix sets
\bea \label{masaop1}
X=\left( \begin{array}{ccc} 0 & \alpha & 0 \\
                            0 & 0 & -\alpha^{T} \\
                            0 & 0 & 0
          \end{array} \right), \quad
K=\left( \begin{array}{ccc} \, & \, & 1 \\
                            \, & I_{\mu} &  \, \\
                             1 & \, & \,
          \end{array} \right),  \nonumber \\
\alpha=(a_{1}, \ldots, a_{\mu}), \quad a_{j} \in  \mathbf{R}. 
\eea
The entries in $\alpha$ are free, and the dimension  of $M$ is hence
\bea
dimM=p-1=\mu
\eea

The algebra $o(2l+1,1)$ has a single (noncompact) Cartan subalgebra, 
corresponding to the orthogonal decomposition $l(2,0) \oplus (1,1)$. 
The algebra $o(2l,1)$ has two inequivalent Cartan subalgebras, 
corresponding to the decompositions $l(2,0) \oplus (0,1)$ (compact) 
and $(1,0) \oplus (1,1) \oplus l(2,0)$ (noncompact).

The situation is illustrated on Fig.1.

\medskip


\setlength{\unitlength}{0.00050000in}%
\begingroup\makeatletter\ifx\SetFigFont\undefined
\def\x#1#2#3#4#5#6#7\relax{\def\x{#1#2#3#4#5#6}}%
\expandafter\x\fmtname xxxxxx\relax \def\y{splain}%
\ifx\x\y   
\gdef\SetFigFont#1#2#3{%
  \ifnum #1<17\tiny\else \ifnum #1<20\small\else
  \ifnum #1<24\normalsize\else \ifnum #1<29\large\else
  \ifnum #1<34\Large\else \ifnum #1<41\LARGE\else
     \huge\fi\fi\fi\fi\fi\fi
  \csname #3\endcsname}%
\else
\gdef\SetFigFont#1#2#3{\begingroup
  \count@#1\relax \ifnum 25<\count@\count@25\fi
  \def\x{\endgroup\@setsize\SetFigFont{#2pt}}%
  \expandafter\x
    \csname \romannumeral\the\count@ pt\expandafter\endcsname
    \csname @\romannumeral\the\count@ pt\endcsname
  \csname #3\endcsname}%
\fi
\fi\endgroup
\begin{picture}(5862,3969)(1576,-5050)
\thicklines
\put(3226,-2011){\line( 0,-1){1125}}
\put(3226,-3136){\line( 0, 1){ 75}}
\put(2401,-3136){\line( 1, 0){825}}
\put(3226,-2011){\line( 1, 0){1575}}
\put(4801,-2011){\line( 0, 1){600}}
\put(4801,-2011){\line( 0,-1){675}}
\put(4801,-2686){\line( 1, 0){2550}}
\put(3226,-4261){\makebox(11.1111,16.6667){\SetFigFont{10}{12}{rm}.}}
\put(3226,-4261){\line( 1, 0){4200}}
\put(4801,-1411){\makebox(11.1111,16.6667){\SetFigFont{10}{12}{rm}.}}
\put(4801,-1411){\line( 1, 0){2550}}
\put(3226,-3136){\makebox(11.1111,16.6667){\SetFigFont{10}{12}{rm}.}}
\put(3226,-3136){\line( 0,-1){1125}}
\put(1576,-3211){\makebox(0,0)[lb]{\smash{\SetFigFont{11}{13.2}{rm}o(p,1)}}}
\put(3751,-1861){\makebox(0,0)[lb]{\smash{\SetFigFont{11}{13.2}{rm}OD}}}
\put(5401,-1261){\makebox(0,0)[lb]{\smash{\SetFigFont{11}{13.2}{rm}l(2,0)+(k,1) \, \, (all $p \geq 2$)}}}
\put(5176,-2536){\makebox(0,0)[lb]{\smash{\SetFigFont{11}{13.2}{rm}l(2,0)+(1,1)+(1,0)\, \, (for p even)}}}
\put(3976,-4111){\makebox(0,0)[lb]{\smash{\SetFigFont{11}{13.2}{rm}OID $\Leftrightarrow$ MANS \, \, (all $p \geq 2$)}}}
\put(1651,-5011){\makebox(0,0)[lb]{\smash{\SetFigFont{9}{11.0}{it}Fig. 1 MASAs of o(p,1)}}}
\end{picture}


\subsection{Behavior of MASAs of e(p,0) under the action of the group 
O(p+1,1)}

\begin{theorem}
All MASAs of e(p,0) inequivalent under E(p,0) are also inequivalent under 
the action of
the group O(p+1,1) and 
are also MASAs of o(p+1,1).
\end{theorem}

{\it Proof}:
A MASA of e(p,0) can be represented in matrix form as follows
\bea
M_e=\left( \begin{array}{cccccc}  M_{1} & \ & \ & \ & 0 \\ 
                                \ & \ddots & \ & \ & \vdots \\
                                \ & \ & M_{l} & \ & 0 \\
                                \ & \ & \ & 0_{k_{+}} & x^{T} \\
                                \ & \ & \ & \ & 0_{1}
          \end{array} \right), \qquad M_i=\left( \begin{array}{cc} 0 & a_i \\
           -a_i & 0 \end{array} \right), \\
i=1, \ldots, l \qquad a_i \in \mR, \qquad          
K_e =\left( \begin{array}{ccc}   I_{2l} & \ & \  \\
                                 \ & I_{k_{+}} & \ \\
                                  \ & \ & 0_{1} 
                \end{array} \right) \nonumber
\eea
which corresponds in o(p+1,1) to the following matrix realization:
\bea
M_e=\left( \begin{array}{cccccc}  M_{1} & \ & \ & \ & \ & 0 \\ 
                                \ & \ddots & \ & \  & & \vdots \\
                                \ & \ & M_{l} & \ & & 0 \\
                                 \ & & \ & 0 & x & 0 \\
                                \ & \ & & \ & 0_{k_{+}} & -x^{T} \\
                                \ & \ & & \ & \ & 0
          \end{array} \right), \\
K_e =\left( \begin{array}{cccc}  I_{2l} & \ & \ & \ \\
                                   \ & \ & \ & 1 \\
                                 \ & \ & I_{k_{+}} & \ \\
                                  \ & 1  & \ & \  
                \end{array} \right) \nonumber
\eea
which is an orthogonally decomposable MASA of o(p+1,1) with decomposition: $l(2,0) 
\oplus {\rm MANS \, of} \,
o(p-2l+1,1)$ (realized as in eq. (\ref{masaop1})).
$\hfill\Box$

\subsection{Summary of MASAs of e(p,0)}

The classification of MASAs of e(p,0) can be summed up in terms of orthogonal 
decompositions of the Euclidean space $M(p,0) \equiv M(p)$.

\begin{theorem} \ \label{th3.3} \\
\begin{zoznamnum}
\item 
Orthogonally indecomposable MASAs exist only for p=1 and p=2. Namely:
\bea
p & = 1   & \{P_1\} \label{3.5} \\
p & = 2   & \{M_{12}\} \label{3.6}
\eea
\item
All MASAs of e(p,0) are obtained by orthogonally decomposing the space M(p) according to 
a pattern
\bea
M(p)=lM(2) \oplus kM(1), \qquad p=2l+k \label{3.7}
\eea
and taking a MASA of type (\ref{3.6}) in each M(2) space and of type (\ref{3.5}) 
in each M(1) space.
\item
For each partition $p=2l+k, 0 \leq l \leq \left[p \over 2 \right]$
we have precisely one conjugacy class of MASAs, both under the isometry group E(p,0)
and the conformal group $O(p+1,1)$.  
\end{zoznamnum}
\end{theorem}



\section {MASA's of e(p,1)}
\setcounter{equation}{0}
\subsection{Splitting MASA's of e(p,1)}
For $e(p,1)$ only the values $k_{-}=0,1$ and $k_{0}=0,1$ are allowed, while 
$0 \leq k_{+} \leq p$. We can write a MASA in the following form
\bea
M(\kp, \km, \kz) \equiv M=\left( \begin{array}{cccccc} M_{0} & \ & \ & \ & \ & \gamma^{T} \\
                               \ & M_{1} & \ & \ & \ & 0 \\ 
                               \ & \ & \ddots & \ & \ & \vdots \\
                               \ & \ & \ & M_{l} & \ & 0 \\
                               \ & \ & \ & \ & 0_{k_{+}} & x^{T} \\
                               \ & \ & \ & \ & \ & 0_{1}
          \end{array} \right) \;,  \label{masa} \\
K_e =\left( \begin{array}{cccc}    K_{0} & \ & \ & \ \\
                                \ & I_{2l} & \ & \ \\
                                 \ & \ & I_{k_{+}} & \ \\
                                  \ & \ & \ & 0_{1} 
                \end{array} \right), \qquad sgnK_0=(p-\kp-2l,1) \nonumber
\eea
where
$M_{i}=\left( \begin{array}{cc} 0 & a_{i} \\ -a_{i} & 0 \end{array} \right) \;,$
$x \in \mathbf{R}^{1 \times k_{+}}$. 
From now on we will only write the form of $M_{0}$, $\gamma$ and $K_{0}$ together
with conditions on the values $l$ and $k_{+}$. The complete MASA can be obtained 
by substituing the appropriate $M_{0}$, $\gamma$ and $K_{0}$  into equation (\ref{masa}). 
We denote the dimensions of these MASAs  $\hbox{dim}M(\kp, \km, \kz) \equiv d. $ 
 
\begin{theorem}
Three different kinds of splitting
MASAs exist. They are characterized by   the triplet $(k_{+}, k_{-}, k_{0})$:
\begin{zoznamAlph} 
\item \, \,  $M(k_{+},1,0)$, $0 \leq \kp \leq p$
\newline 
\bea M_{0}=0 \in {\mathbf{R}}, \quad \gamma^{T}=z \in {\mathbf{R}} \quad  and
\quad  K_{0}=-1 \label{p1spl:k10}
\eea
$p-k_{+}$ is even,  $0  \leq l \leq \frac{p-k_{+}}{2}, d=dim M(\kp,1,0)=1+l+\kp, \left[ p+3 \over 2 \right] \leq d \leq p+1$
\newline
\item \, \, $M(k_{+} , 0 , 0 ), 0 \leq \kp \leq p-1 $   
\bea M_{0}=\left( \begin{array}{cc} c & 0 \\
                                    0 & -c  
                    \end{array} \right), \qquad
\gamma^{T}=\left( \begin{array} {c}  0 \\ 0 
                  \end{array} \right), \qquad                     
K_{0}=\left( \begin{array}{cc}   0 & 1   \\
                                 1 & 0     
              \end{array} \right) \label{p1spl:k00}
\eea
\newline
where $p-k_{+}$ is odd, $0  \leq l \leq  {p-k_{+}-1 \over 2}$,
 $d = dimM(\kp,0,0) =1+l+\kp, \left[ {p+2 \over 2}\right] \leq d \leq p$ \newline

\item \, \,  $M(k_{+} , 0 , 1 )$, $0 \leq \kp \leq p-2$ 
\bea M_{0}=\left( \begin{array}{ccc} 0 & \alpha & 0 \\
                                      0 & 0 & -\alpha^{T} \\
                                      0 & 0 & 0                                 
               \end{array} \right), \qquad
\gamma^{T}=\left( \begin{array} {c} z\\ 0_{\mu} \\ 0 
                  \end{array} \right), \qquad                
K_{0}=\left( \begin{array}{cccccc}  \ & \ & 1  \\
                                     \ & I_{\mu} & \  \\
                                          1 & \ & \ 
                   \end{array} \right) \label{p1spl:k01}
\eea
where $1 \leq \mu \leq p-1$ and $0 \leq l \leq {{p-k_{+}-2} \over 2}, z \in {\bf R}$, $\alpha \in 
{\mathbf{R}}^{1 \times \mu}, d=dimM(\kp,0,1)=\mu+l+\kp+1, \left[ p+3 \over 2 \right] \leq d \leq p$.
\end{zoznamAlph}
All entries $a_i, x, z, \alpha$ and $c$ are free. 
\label{tp1spl}
\end{theorem}

{\it Proof.}\ Let us use the representation (\ref{H}) 
of $e(p,1).$  The translations are represented by the matrix 
$Y$ with $X = 0.$ We run 
through the three translation subalgebras $T$ fixed in  Theorem \ref{tp1spl}
and for each of them find their centralizer $C(T)$ in $o(p,1),$
i.e. the set of matrices $X$ and $Y,$ such that we have 
\begin{equation} 
[Y(X,0), Y(0, \alpha)] = 0 \label{4.5}
\end{equation} 
for the chosen set of the translations $\alpha.$ 
We must then determine all MASAs of $C(T)$ such that they 
commute only with $T$ and with no other translations.

\begin{zoznamAlph}
\item For $T = T(\kp, 1, 0)$ we have $
C(T) \sim o(p-\kp,0)$ which has only one MASA: the Cartan subalgebra. 
The condition 
$p-\kp$ being even is needed, otherwise the MASA 
will commute with $\kp+1$ positive 
length 
vectors. We thus arrive at  eq.(\ref{p1spl:k10}). 

\item For $T= T(\kp, 0, 0)$ we obtain 
$C(T) \sim o(p-\kp, 1).$ The MASAs of $o(p-\kp, 1)$ 
are known (see section 3.2 above and also \cite{verop}). 
Any MASA of $o(p-\kp, 1)$ containing a nilpotent element 
will also commute with an isotropic vector in $T,$ not 
contained in $T(\kp, 0, 0).$ Hence we need only 
to consider a Cartan subalgebra of $o(p-\kp, 1).$ 
Moreover, it must be noncompact, or it will 
commute with a negative length vector in $T.$ 
Finally, if $p-\kp$ is even, the MASA will commute with $\kp+1$ 
positive length vectors in  $T.$ We arrive 
at the result in eq.(\ref{p1spl:k00}).

\item Take $T = T(\kp, 0, 1).$ We obtain 
$C(T) \sim e(p-\kp-1, 0),$ an Euclidean Lie algebra realized as a subalgebra of 
$o(p-\kp, 1),$ e.g. by the matrices 
\beq
Z = \left( \begin{array}{ccc}
0 & \nu & 0 \\
0 & R & -\nu^T \\ 
0 & 0 & 0 \\
\end{array} \right) \;, \label{4.6}
\eeq
where  $R+R^T = 0, R\in {\hbox{\bf R}}^{(p-\kp-1)\times(p-\kp-1)}, 
\nu\in {\hbox{\bf R}}^{1\times (p-\kp-1)}.$ 
\end{zoznamAlph}

Applying the Theorem \ref{ep0} we obtain the result given 
in eq.(\ref{p1spl:k01}). The results concerning the dimensions of the 
MASAs are obvious;
they amount to counting the number of free parameters in $M_0, M_i, \gamma$ 
and $x$ in the matrix (\ref{masa}).
$\hfill\Box$


\subsection{Nonsplitting MASA's of e(p,1)}
First we describe the general procedure for finding nonsplitting MASAs of $e(p,q)$.

Every nonsplitting MASA $M(\kp, \km, \kz)$ of $e(p,q)$ is obtained from a splitting 
one by the following procedure: 

\begin{zoznamnum}
\item Choose a basis for $C(\kp, \km, \kz)$ and $T(\kp, \km, \kz)$ e.g. 
$C(\kp, \km, \kz) \sim \\
\{B_{1}, \ldots , B_{J}\}$, $T(\kp, \km, \kz) \sim \{X_{1}, 
\ldots X_{L}\}$.
\item  Complement the basis of $T(\kp, \km, \kz)$ to a basis of $T(n)$.
\begin{eqnarray*}
T(n) / T(\kp, \km, \kz) = \{Y_{1}, \ldots , Y_{N}\}, \qquad L+N=n.
\end{eqnarray*}
\item Form the elements 
\beq
\tilde B_{a}= B_{a} + \sum_{j=1}^{N} \tilde \alpha_{aj} Y_j, \qquad a=1, \ldots, J
\eeq
where the constants $\tilde \alpha_{aj}$ are such that $\tilde B_{a}$ form an 
Abelian Lie algebra
$[\tilde B_{a},\tilde B_{b}]=0$. This provides a set of linear equations for the 
coefficients $\tilde \alpha_{aj}$. Solutions  $\tilde \alpha_{aj}$ are called 
1-cocycles and they provide the Abelian subalgebras $\tilde M (\kp, \km, \kz) \sim 
\{\tilde B_{a}, X_b\} \subset e(p,q)$.
\item Classify the subalgebras $\tilde M (\kp, \km, \kz)$ into conjugacy classes 
under the action of the group $E(p,q)$. This can be done in two steps.
\begin{zoznamrom}
\item Generate trivial cocycles $t_{aj}$, called coboundaries, using the translation group $T(n)$
\beq
e^{p_jP_j}\tilde B_{a}e^{-p_jP_j}=\tilde B_{a}+p_j[P_j,\tilde B_{a}]=\tilde B_{a}+\sum_jt_{aj}P_j.
\eeq
The coboundaries should be removed from the set of the cocycles. If we have $\tilde 
\alpha_{aj} =t_{aj}$ for all $(a,j)$ the algebra is splitting (i.e. equivalent to a splitting one).
\item Use the normalizer of the splitting subalgebra in the group $O(p,q)$ to further 
simplify and classify the nontrivial cocycles.

\end{zoznamrom}
\end{zoznamnum}

\begin{theorem}
Nonsplitting MASA's of $e(p,1)$ are obtained from splitting ones of type $C$ in  Theorem~\ref{tp1spl} and
are conjugate to  precisely one MASA of the  form  \\
i) for $\mu \geq 2$
\bea M_{0}=\left( \begin{array}{ccc} 
                      0 & \alpha & 0  \\
                      0 & 0 & -\alpha^{T}  \\
                      0 & 0 & 0  \\
                   \end{array} \right), \qquad
\gamma^{T}=\left( \begin{array} {c} z\\ A\alpha^{T} \\ 0
                  \end{array} \right)                   
                    \label{p1nspl}
\eea	
where A is a diagonal matrix with  $a_1=1 \geq |a_2| \geq \ldots \geq |a_\mu| \geq 0$  and $TrA=0$,
$K_{0}$ is as in (\protect\ref{p1spl:k01}) 
\newline
ii)\, \,  for $\mu =1$ we have a special case for which the 
nonsplitting MASA has the following form
\bea M_{0}=\left( \begin{array}{ccc} 0 & a & 0  \\
                                      0 & 0 & -a  \\
                                      0 & 0 & 0  \\
                  \end{array} \right),  \, \, \, \, \,
\gamma^{T}=\left( \begin{array} {c} z\\ 0 \\ a
                  \end{array} \right), \, \, \, \, \,                  
K_{0}=\left( \begin{array}{ccc} 0 & 0 & 1  \\
                     		       0 & 1 & 0  \\
 				       1 & 0 & 0  \\
              \end{array} \right). \label{p1nspl-m}
\eea 
No other nonsplitting MASAs of $e(p,1)$ exist.
\label{tp1:nspl}
\end{theorem}

{\it Proof}:
The nonsplitting MASA is represented in general as follows
\bea Z_{e}=\left( \begin{array}{cccccc}  M_{0} & \, & \, & \, & \, & \beta_{0}^T \\
                                        \, &  M_{1} & \, & \, & \, & \beta_{1}^T \\  
                                          \, &   \, & \ddots & \, & \, & \vdots \\
                                         \, & \, & \, & M_{l} & \,  & \beta_{l}^T \\ 
                                          \, & \, & \, & \, & 0_{k_{+}} & x^{T} \\ 
                                          \, & \, & \, & \, & \, & 0_{1}
                  \end{array} \right)
\eea 
where $\beta_{0} \in R^{1 \times (p-\kp-2l)}$ and $\beta_{i} \in {\mathbf R}^{1 \times 2}, i =1, \ldots l$, depend linearly on the free 
entries in the MASA of $o(p,1)$ {\it i.e.} the matrices $M_{i}, 0 \leq i \leq l$.
We impose the commutativity $[Z_{e},Z_{e}']= 0$ and obtain
\bea M_{i}\beta_{i}^{'T}= M'_{i}\beta_{i}^T, \qquad i=0, \ldots, l \label{mibeta}.
\eea
From eq.(\ref{mibeta}) we see that vectors $\beta_{i}$ depends linearly on the 
matrices $M_{i}$ only. The block $(M_{i},\beta_{i})$, $\beta_{i}=(a_i,a_{i+1})$ 
for $i=1, \ldots, l$ 
represents elements of the type
$$ L_{i,i+1}+a_{i}P_{i}+a_{i+1}P_{i+1}, \qquad 1 \leq i \leq p.$$ 
In all cases the coefficients $a_{i}$ are coboundaries, since we have 
\bea 
\exp{(\alpha_{i}P_{i}+\alpha_{i+1}P_{i+1})} L_{i,i+1} \exp{(-\alpha_{i}
P_{i}-\alpha_{i+1}P_{i+1})}= \nonumber \\
L_{i,i+1} + \alpha_{i}P_{i+1}-\alpha_{i+1}P_{i}.
\eea
The coefficients $\alpha_{i}$ can be chosen so as to annul $a_{i}$ 
and $a_{i+1}$.
Thus we have 
\bea \beta_{j}=0, \qquad 1 \leq j \leq l \eea
for all nonsplitting MASAs of $e(p,1)$. Hence for case $A)$ from 
Theorem~\protect\ref{tp1spl} there are no nonsplitting MASAs. 
In the case $B)$ the block  $(M_{0},\beta_{0})$ represents the element 
of the type $ L_{p,p+1}+a_{p}P_{p}+a_{p+1}P_{p+1}$.
Here again the coefficients $a_{i}$ are coboundaries, since we have 
\bea \exp{(\alpha_{p}P_{p}+\alpha_{p+1}P_{p+1})} L_{p,p+1} 
\exp{(-\alpha_{p}P_{p}-\alpha_{p+1}P_{p+1})}= \nonumber \\
L_{p,p+1} + \alpha_{p}P_{p+1}+\alpha_{p+1}P_{p}
\eea
and the coefficients $\alpha_{i}$ can be chosen so as to annul $a_{p}$ 
and $a_{p+1}$. We have that $\beta_{0}=0$, and there are no nonsplitting MASAs.
In the case  $C)$ the nonsplitting part of $M_{0}$ is as follows
\bea Z_{0}=\left( \begin{array}{cccc} 0 & \alpha & 0 & 0 \\
                                      0 & 0 & -\alpha^{T} & \beta_0^{T} \\
                                      0 & 0 & 0 & y \\
                                       0 & 0 & 0 & 0_{1} 
                    \end{array} \right).
\eea
Commutativity  $[Z_{e},Z_{e}']= 0$ gives us the following conditions
\bea 
\alpha \beta_{0}^{'T}&=& \alpha' \beta_{0}^{T}, \\
\alpha^{T} y'&=& \alpha^{'T} y, \qquad y \in {\mathbf R}
\eea
which gives 
\bea \beta_{0}^{T}= A \alpha^{T}, \\
      y=\mu \alpha^{T} \label{muy}
\eea
where $A$ is a matrix and $\mu$ is a row vector.
 
Looking again at the commutativity condition with eq.(\ref{muy}) satisfied, 
we find that 
\bea A=A^{T} \qquad   {\rm and} \qquad \mu=0.
\eea
The symmetric matrix $A$ represents the 1-cocycles. The coboundaries are 
represented by the matrix $\delta I$ and we use them to set $Tr A=0$. To 
further simplify and classify we use the normalizer of 
the splitting MASA in the group $o(p,1)$. The normalizer is represented by block 
diagonal matrices  of the same block structure as in eq.(\ref{masa}). The part 
acting on $M_{0}$ is represented by
\bea G=diag(g,G_{0},g^{-1},1), \qquad {\rm satisfying} \qquad G_{0}G_{0}^{T}=I. 
\eea
Computing 
\bea GM_{0}G^{-1}=M_{0}' \eea
gives the following transformation of $A$
\bea
A'={1 \over g} (G_0 A G_{0}^{T}).
\eea
We use the matrix $G_0$ to diagonalize $A$ and to order the eigenvalues. 
The normalization $a_1=1$ is due to a choice of $g$.
The proof of the case $ii)$ is almost identical to the previous one and we 
omit it here.
The dimension of the 
nonsplitting subalgebra is the same as the dimension of the corresponding 
splitting subalgebra.
$\hfill\Box$

\subsection{A decomposition theorem for MASAs of e(p,1)}

Again, all the results of this section can be summed up in a decomposition theorem.
\begin{theorem} \ \label{th4.3} \\
\begin{zoznamnum}
\item
Indecomposable MASAs of e(p,1) exist for all values of p, namely 
\bea
p=0  &:&  \{P_0\} \label{4.25} \\
p=1  &:&  \{L_{01}\} \label{4.26} \\
p=2  &:&  \{P_0-P_1, L_{02}-L_{12}+\kappa(P_0+P_1)\}, \label{4.27}\\
     & & \kappa=0,\pm 1 \nonumber \\
p \geq 3  &:&  \{P_0-P_1, L_{0j}-L_{1j}+ a_jP_j)\},  \label{4.28} \\
     & & j=2, \ldots p, a_2=1 \geq |a_3| \geq \ldots \geq |a_p| \geq 0, \nonumber \\
     & & \sum a_i = 0 \nonumber \\
or   & &  a_2=a_3=\ldots=a_p=0 .\nonumber
\eea
MASAs corresponding to different values of $\kappa$, or different sets 
$(a_2, \ldots a_p)$ 
are mutually inequivalent under the connected component of $E(p,1)$. If the 
entire group  $E(p,1)$ is allowed (containing $O(p,1)$, rather than only 
$SO(p,1)$), then $\kappa=-1$ is equivalent to $\kappa=1$ and can be omitted.
\item
All MASAs of e(p,1) are obtained by orthogonally decomposing the Minkowski space M(p,1) 
according to the pattern
\bea \label{4.29}
M(p,1) = M(k,1) \oplus lM(2,0)  \oplus mM(1,0), \qquad p=k+2l+m, \\
0 \leq k \leq p, \qquad
0 \leq l \leq \left[p \over 2 \right] \nonumber 
\eea
and taking a MASA of type (\ref{3.5}) for each M(1), of type (\ref{3.6}) for each 
M(2) and of type (\ref{4.25}),(\ref{4.26}),(\ref{4.27}) or (\ref{4.28}) for M(k,1).
\item
Each decomposition (\ref{4.29}) and each choice of constants $\kappa$ and $\{a_j\}$, 
respectively, provides a different MASA (mutually inequivalent under the group E(p,1)).
\end{zoznamnum}
\end{theorem}



\section {Embedding of MASAs of e(p,1) into the conformal algebra o(p+1,2)}
\setcounter{equation}{0}
\subsection{Introductory comments}

Let us realize the algebra $o(r,2)$ by matrices $X$ satisfying
\bea
XK+KX^T=0, \qquad K,X \in \mR, \qquad K=K^T, \qquad sgnK=(r,2).
\eea

A MASA of $o(r,2)$ will be called {\it orthogonally decomposable} (OD) if all 
matrices representing the MASA can be simultaneously transformed by some matrix 
$G$, together with the matrix $K$, into block diagonal sets of the form
\bea
\tilde X=\left( \begin{array}{cccc} X_1 & \, & \, & \, \\
                                     \, & X_2 & \, & \, \\
                                     \, & \, & \ddots & \, \\
                                     \, & \, & \, & X_j
                 \end{array} \right), \qquad
\tilde K=\left( \begin{array}{cccc} K_1 & \, & \, & \, \\
                                     \, & K_2 & \, & \, \\
                                     \, & \, & \ddots & \, \\
                                     \, & \, & \, & K_j
                 \end{array} \right), \label{5.2} \\
\tilde X=GXG^{-1}, \qquad \tilde K=GKG^{T}, \qquad G \in GL(r+2,\mR). \nonumber
\eea

If no such matrix $G$ exists, the MASA is {\it orthogonally indecomposable} (OID).

A MASA can be orthogonally indecomposable, but {\it not absolutely indecomposable} 
(OID, but NAOID). This means it is orthogonally decomposable after complexification 
of the ground field.

Let us now present some results on MASAs of $o(r,2)$ that can be extracted from 
Ref.~\cite{verop}.


\subsection{MASAs of o(r,2)}
We shall first consider $r \geq 3$, then treat the case $r=2$ separately.

\begin{proposition} \label{prop5.1}
Precisely 3 types of MASAs exist for $r=2k \geq 4$, 2 for $r=2k+1 \geq 3$.
\begin{zoznamnum}
\item Orthogonally decomposable MASAs (any r)
\item Absolutely orthogonally indecomposable MASAs (any r)
\item Orthogonaly indecomposable, but not absolutely orthogonally indecomposable MASAs (r=2k)
\end{zoznamnum}
\end{proposition}

\begin{proposition} \label{prop5.2}
Every orthogonally decomposable MASA of o(r,2) can be represented
in the form (\ref{5.2}) where each $\{X_i, K_i\}$ represents an orthogonally 
indecomposable MASA of lower dimension. The allowed decomposition patterns are
\begin{zoznamnum}
\item $(r,2)=(s,2)+l(2,0), \qquad r=s+2l, \qquad  l\geq 1$
\item $(r,2)=(s,2) + (1,1) +l(2,0), \qquad r=s+2l+1$.
\end{zoznamnum}
\end{proposition}

A {\it maximal Abelian nilpotent subalgebra} (MANS) of $o\pq$ is characterized 
by its Kravchuk signature $(\lambda \mu \lambda)$, a triplet of nonnegative integers 
satisfying
\bea 
2\lambda+\mu=p+q, \qquad \mu \geq 0, \qquad 1 \leq \lambda \leq q \leq p.
\eea

For a given MANS $M$ the positive integer $\lambda$ is the dimension of the kernel 
of $M$ and also the codimension of the image space of $M$. For a given signature 
$(\lambda \, \mu \lambda)$ the MANS $M$ can be transformed into Kravchuk 
normal form, namely
\bea
X=\left( \begin{array}{ccc} 0 & A & Y \\
                             0 & S & -K_{0}A^{T} \\
                             0 & 0 & 0
           \end{array} \right), \quad
K=\left( \begin{array}{ccc} \, & \, & I_{\lambda} \\
                             \, & K_{0} & \, \\
                             I_{\lambda} & \, & \,
           \end{array} \right), \nonumber \\
A \in \mathbf{R^{\lambda \times \mu}}, \qquad
Y=-Y^{T} \in \mathbf{R^{\lambda \times \lambda}}, \qquad
SK_{0}+K_{0}S^{T}=0, \nonumber \\
S \in \mathbf{R^{\mu \times \mu}}, \qquad
K_{0}=K_{0}^{T} \in \mathbf{R^{\mu \times \mu}}, \qquad
sgn K_{0}=(p-\lambda, q-\lambda). \label{KS}
\eea
The matrix $S$ is nilpotent, the matrix $K_{0}$ is fixed. The classification 
of MANSs of $o(p,q)$ reduces to a classification of matrices $A$, $S$ and $Y$ 
satisfying the commutativity relation $[X,X']=0$:

\bea
AK_{0}A^{'T}=A'K_{0}A^{T}, \qquad AS'=A'S, \qquad [S,S']=0.
\eea

Two types of MANSs of $o(p,q)$ exists
\begin{zoznamnum}
\item
{\it Free-rowed MANS}. There exists a linear combination of the $\lambda$ rows 
of the matrix $A$ in (\ref{KS}) that contains $\mu$ free real entires.
\item
{\it Non-free-rowed MANS}. No linear combination of the $\lambda$ rows of $A$ contains more than $\mu - 1$ real free entries.
\end{zoznamnum}

\begin{proposition} \label{prop5.3}
An absolutely orthogonally indecomposable MASA of o(r,2) is a MANS. Three types 
of MANSs of o(r,2) exists. Using the metric 
\bea
K=\left( \begin{array}{ccc} \, & \, & 1 \\
                             \, & K_0 & \,  \\
                            1 & \, & \, 
         \end{array} \right), \quad
K_0=\left( \begin{array}{ccc} \, & \, & 1 \\
                             \, & I_{r-2} & \,  \\
                            1 & \, & \, 
         \end{array} \right) \label{5.6}
\eea
they can be written as
\begin{zoznamnum}
\item Kravchuk signature (1 r 1), free rowed
\bea
X=\left( \begin{array}{ccc} 0 & \alpha & 0 \\
                             0 & 0 & -K_0\alpha^T   \\
                           0 & 0 & 0 
         \end{array} \right), \qquad \alpha \in \mR^{1 \times r} \label{5.7}
\eea
\item Kravchuk signature (1 r 1), non-free rowed
\bea
X=\left( \begin{array}{cccccc} 0 & a & \alpha & 0 & b & 0 \\
                               \, & 0 & 0 & a & 0 & -b \\
                              \, & \, & 0 & 0 & 0 & -\alpha^{T} \\
                               \, & \, & \, & 0 & -a & 0 \\
                               \, & \, & \, & \, & 0 & -a \\
                                \, & \, & \, & \, & \, & 0 
           \end{array} \right), \label{5.8} \\
a,b \in \mR, \qquad \alpha \in \mR^{1 \times (r-3)} \nonumber
\eea
\item Kravchuk signature (2 r-2 2), free rowed
\bea
X=\left( \begin{array}{ccccc} 0 & 0 & \alpha & x & 0 \\
                              0 & 0 & \alpha Q & 0 & -x \\
                              \, & \, & \, & -Q\alpha^T & -\alpha^T \\
                              \, & \, & \, & 0 & 0 \\
                               \, & \, & \, & 0 & 0 \\
          \end{array} \right), \label{5.9} \\
\alpha \in \mR^{1 \times (r-2)}, \qquad Q=diag(q_1, \ldots ,q_{r-2}) \neq 0, 
\qquad \sum_{j=1}^{r-2}q_j=0, \nonumber \\
1=q_1 \geq |q_2| \geq \ldots \geq |q_{r-2}| \geq 0. \nonumber
\eea 
\end{zoznamnum} 
\end{proposition}

\begin{proposition} \label{prop5.4}
The algebra $o(2k,2), k \geq 2$ has precisely one class of orthogonally 
indecomposable, but not absolutely indecomposable MASAs. It can be represented by the set of matrices $\{X,K\}$
\bea
X=\left( \begin{array}{cccccccccc}
           0 & a & b_1 & b_1 & \, & \, & b_{k-1} & b_{k-1} & 0 & c \\
           -a & 0  & b_1 & -b_1 & \, & \, & b_{k-1} & -b_{k-1} & -c & 0 \\
           \, & \, & 0 & a & \, & \, & \, & \, & -b_1 & -b_1 \\
           \, & \, & -a & 0 & \, & \, & \, & \, & -b_1 & b_1 \\
           \, & \, & \, & \, & \, & \, & \, & \, & \, & \, \\
           \, & \, & \, & \, & \, & \, & \, & \, & \, & \, \\
           \, & \, & \, & \, & \, & \, & 0 & a & -b_{k-1} & -b_{k-1} \\
           \, & \, & \, & \, & \, & \, & -a & 0 & -b_{k-1} & b_{k-1} \\
           \, & \, & \, & \, & \, & \, & \, & \, & 0 & a \\
           \, & \, & \, & \, & \, & \, & \, & \, & -a & 0 
          \end{array} \right), \label{5.10} \\
K=\left( \begin{array}{ccccc} & & & 1 &  \\
                              & & &  & 1 \\
                              & & I_{2k-2} & & \\
                              1 &  & & & \\
                               & 1 & & &
         \end{array} \right)
\eea 
\end{proposition}

The algebra $o(2,2)$ is exceptional for two reasons, namely we have $p=q=$even and
moreover, it is semisimple rather than simple. Two orthogonal decompositions exists, 
namely those of Proposition ~\ref{prop5.2} with $s=0, l=1$ in the first case, 
$s=1, l=0$ in the second. The MANS of eq.(\ref{5.7}) also exist in this case, as 
does the MASA (\ref{5.10}), not however (\ref{5.8}) and (\ref{5.9}). On the other 
hand, two further MASAs exist, both decomposable, but not orthogonally decomposable. 
In terms of matrices, they are represented by
\bea 
X=\left( \begin{array}{cccc} a & b & \ & \ \\
                             \ & a & \ & \ \\
                             \ & \ & -a & -b  \\
                             \ & \ & \ & -a 
        \end{array} \right), \qquad
K= \left( \begin{array}{cc} 0 & J \\
                             J & 0 
         \end{array} \right), \qquad
J= \left( \begin{array}{cc} 0 & 1 \\
                             1 & 0 
         \end{array} \right) \label{5.11}       
\eea
and
\bea
X=\left( \begin{array}{cccc} a & b & \, & \, \\
                             -b & a & \, & \, \\
                             \, & \, & -a & -b  \\
                             \, & \, & b & -a 
        \end{array} \right), \qquad
K= \left( \begin{array}{cc} 0 & J \\
                             J & 0 
         \end{array} \right), \label{5.12}
\eea
respectively. Thus $o(2,2)$ has 6 classes of MASAs.
Propositions \ref{prop5.1},\ldots ,\ref{prop5.4} as well as the results for 
$o(2,2)$, are proven in Ref~\cite{verop}.

Let us now sum up the results on MASAs of $o(p,2)$ in terms of the "physical" 
basis (\ref{diffop}), (\ref{conf}), starting from orthogonally indecomposable ones.
\begin{zoznamnum}
\item
The MANS (\ref{5.7}) of $o(r,2)$ corresponds to the translations
\bea
\{P_0,P_1, \ldots P_{r-1}\} \label{5.13}
\eea
and is contained in $e(r-1,1)$.
\item The MANS (\ref{5.8}) of $o(r,2)$ corresponds to
\bea
\{P_0-P_1,L_{02}-L_{12}+P_0+P_1,P_3, \ldots P_{r-1}\} \label{5.14}
\eea
and is contained in $e(r-1,1)$.
\item The MANS (\ref{5.9}) of $o(r,2)$ corresponds to
\bea
\{P_0-P_1, P_k+q_k(L_{0k}-L_{1k}), k=2, \ldots r-1\} \label{5.15}
\eea
and is contained in $e(r-1,1)$.
\item The MANS (\ref{5.10}) of $o(2k,2)$ corresponds to
\bea
\{2(L_{23}+L_{45}+\ldots +L_{2k-2,2k-1})+(P_0-P_1)-(C_0+C_1), \nonumber \\
P_j+P_{j+1}+L_{0j}+L_{1j}-L_{0,j+1}-
L_{1,j+1}, j=2, \ldots 2k-2, P_0+P_1\} \label{5.16}
\eea
and is not contained in $e(r-1,1)$.
\item For $o(2,2)$ case (\ref{5.11}) corresponds to
\bea
\{P_0-P_1,D-L_{01}\} \label{5.17}
\eea
and (\ref{5.12}) to
\bea
\{D-L_{01},P_0-P_1+(C_0+C_1)\}. \label{5.18}
\eea
They are not contained in $e(1,1)$.
\end{zoznamnum}
In the orthogonally decomposable MASAs each component is an orthogonally 
indecomposable MASA of one of the types listed above.


\subsection{MASAs of e(p,1) classified under the group O(p+1,2)}

Let us make use of the realization (\ref{2.25}) of the algebra
$o(p+1,2)$ and choose $K_0$ as in eq.(\ref{p1spl:k01}). The algebra
$e(p,1) \subset o(p+1,2)$ is represented as follows:
\bea \label {5.19}
X=\left( \begin{array}{ccccc} 0 & p_{+} & \alpha & p_{-} & 0 \\
                              0 & k & \beta & 0 & -p_{-} \\
                              0 & -\gamma^T & R & -\beta^T & -\alpha^T \\
                              0 & 0 & \gamma & -k & 0 \\
                              0 & 0 & 0 & -p_{+} & 0
          \end{array} \right), \\
p_-,p_+,k \in \mR, \qquad \alpha,\beta,\gamma \in \mR^{1 \times (p-1)},
\qquad R=-R^T \in \mR^{(p-1) \times (p-1)}. \nonumber
\eea
In eq.(\ref{5.19}) $R$ represents rotations in the subspace $\mR^{p-1}$
and further we have
\bea
p_- \sim P_0-P_1, \qquad p_+ \sim P_0+P_1, \qquad \alpha \sim (P_2,
\ldots, P_k), \nonumber \\
k \sim L_{01}, \qquad \beta \sim (L_{02}-L_{12}, \ldots
,L_{0p}-L_{1p}), \\
\gamma \sim (L_{02}+L_{12}, \ldots
,L_{0p}+L_{1p}). \nonumber
\eea 

We shall use a transformation represented by a matrix $G \in O(p,2)$,
$G \in E(p,1)$, namely
\bea \label{5.21}
G=\left( \begin{array}{ccc} G_0 & & \\
                            & I_{p-1} & \\
                            & & G_0
         \end{array} \right), \qquad
GXG^{-1}=X^{'}, \qquad GKG^T=K.
\eea
The transformation (\ref{5.21}) with $G_0=\left( \begin{array}{cc}0&1
\\1&0 \end{array}\right)$ leaves $R$ and $P_0-P_1$ invariant,
interchanges $\alpha$ and $\beta$, i.e.$P_j$ and $L_{0j}-L_{1j}$ $(j=2,
\ldots p)$  and takes $L_{01}, P_0+P_1$ and $L_{0j}+L_{1j}$ out of the
$o(p,1)$ subalgebra that we will use to conjugate different MASAs of
$e(p,1)$ that are inequivalent under $E(p,1)$.

Let us now consider the individual decompositions of the space
$M(p,1)$ listed in eq.~(\ref{4.29}) of
Theorem ~\ref{th4.3}.

First of all we note that the presence of $o(2)$ subalgebras acting
in the $M(2,0)$ subspaces (for $l \geq 1)$ implies  an orthogonal
decomposition of the corresponding MASA of $o(p+1,2)$. We are then
dealing with Abelian subalgebras (ASA) of the form
\bea
ASA[o(p+1,2)]=l[o(2)] \oplus ASA[o(j+1,2)], \qquad j+2l=p. \label{5.22}
\eea

From now on we only need to consider subalgebras of $e(j,1) \subset
o(j+1,2)$ and their possible conjugacy under $O(j+1,2)$. These MASAs
of $o(j+1,2)$ contain no rotations $L_{ik}$. The following situations
arise.

\begin{zoznamnum}
\item
$k=0, m=p-2l$ in eq.~(\ref{4.29}) and $j=m$. The MASA of $e(j,1)$
consists of translations only:$\{P_0,P_1, \ldots P_j\}$. This is the free rowed 
MANS of $o(j+1,2)$ with Kravchuk signature {\it (1 j+1 1)} as in eq.~(\ref{5.7}) 
and  (\ref{5.13}).
\item 
$k=1, m=p-2l-1$ in eq.~(\ref{4.29}) and $j=m+1$. The MASA of $e(j,1)$ is an orthogonally 
decomposable MASA of $o(j+1,2)$ of the form
\bea
MASA[o(j+1,2)]=o(1,1) \oplus MANS[o(j,1)] \nonumber
\eea
where the MANS of $o(j,1)$ has Kravchuk signature {\it (1 j-1 1)} as
in eq.~(\ref{masaop1}). In the physical basis it is $\{L_{01},P_2,
\ldots ,P_j\}$.
\item
$k=2,m=p-2l-2$ in eq.~(\ref{4.29}) and $j=m+2$, $\kappa \not= 0$ in
eq.~(\ref{4.27}). We have the MASA $\{L_{02}-L_{12} \pm (P_0+P_1),
P_0-P_1, P_2, \ldots, P_j\}$.  This is a non-free-rowed MANS of
$o(j+1,2)$ with Kravchuk signature {\it (1 j+1 1)} as in eq.~(\ref{5.8}) and (\ref{5.14}).
\item
$k=2, m=p-2l-2 $ in eq.~(\ref{4.29}) and $j=m+2$, $\kappa = 0$ in
eq.~(\ref{4.27}). We have the MASA $\{L_{02}-L_{12}, P_{0}-P_{1}, P_3,
\ldots P_j\}$. The transformation (\ref{5.21}) takes this algebra into
$\{P_0-P_1, P_2,L_{03}-L_{13}, \ldots L_{0j}-L_{1j}\}$. Thus, if we
are interested in conformally inequivalent MASAs, we must impose, for
$\kappa \not= 0$, $j \geq 3$, i.e $m \geq 1$ in eq.~(\ref{4.29}). This
MASA is a free rowed MANS of $o(j+1,2)$ with Kravchuk signature {\it
(2 j-2 2)} as in eq.~(\ref{5.9}) and (\ref{5.15}).

\item
$k \geq 3, m=p-2l-k$ in eq.~(\ref{4.29}) and $j=m+k$, $a_2=a_3=\ldots
=a_j=0$ in eq.~(\ref{4.28}). The MASA is $\{P_0-P_1,L_{02}-L_{12},
\ldots L_{0k}-L_{1k}, P_{k+1}, \ldots P_j\}$ and is conformally
equivalent to $\{P_0-P_1,P_2, \ldots P_k, L_{0,k+1}-L_{1,k+1}, \ldots
L_{0j}-L_{1j}\}$. It is a free rowed MANS of $o(j+1,2)$ with Kravchuk
signature {\it (2 j-1 2)} as in eq.~(\ref{5.9}) and (\ref{5.15}).

\item
$k \geq 3, m=p-2l-k$ in eq.~(\ref{4.29}) so  $j=m+k$,
$|a_2|=1 \geq |a_3| \geq \ldots
\geq |a_j|$ in eq.~(\ref{4.28}). The MASA is $\{P_0-P_1,L_{02}-L_{12}+a_{2}P_{2},
\ldots L_{0k}-L_{1k}+a_{k}P_{k}, P_{k+1}, \ldots P_j\}$. Again we have a
free rowed MANS of $o(j+1,2)$ with Kravchuk signature {\it
(2 j-1 2)} as in eq.~(\ref{5.9}) and (\ref{5.15}).
\end{zoznamnum}

We see that the MASAs listed above in cases 4, 5 and 6 are all
related. Indeed, let us fix some value of $j$ and consider the MANS
(\ref{5.9}) of $o(j+1,2)$. Cases 4 and 5 corresponds to the first two
rows in eq.~(\ref{5.9}) being
\bea \label{5.23}
\left( \begin{array}{ccccc} 0 & 0 & \alpha & x & 0 \\
                             0 & 0 & \beta & 0 & -x \end{array}
\right) =\left( \begin{array}{cccccccccc} 0 & 0 & \alpha_2 & \ldots &
\alpha_k & 0 & \ldots & 0 & x & 0 \\
0 & 0 & 0 & \ldots & 0 & \beta_{k+1} & \ldots & \beta_j & 0 & -x 
\end{array} \right)
\eea

The transformation (\ref{5.21}) with 
\bea \label{5.24}
G_0= \left( \begin{array}{cc} 1&1 \\a & b \end{array}\right)
\eea
takes (\ref{5.23}) into the standard form with
\bea \label{5.25}
\left( \begin{array}{c} \alpha \\ \beta \end{array} \right) = \left(
\begin{array}{cccccc}
\alpha_2 & \ldots & \alpha_k & \beta_{k+1} & \ldots & \beta_j \\
a\alpha_2 & \ldots & a\alpha_k & b\beta_{k+1} & \ldots & b\beta_j
\end{array} \right)
\eea 
with $j-1$ free entries in row 1 and $Q=diag(aI_{k-1},bI_{j-k})$, with
\bea \label{5.26}
(k-1)a+(j-k)b=0, \qquad b \not = a 
\eea

An exception occurs when $m=0$. The algebra then is $\{P_0-P_1,
L_{02}-L_{12}, \ldots L_{0j}-L_{1j}\}$. This is equivalent to
$\{P_0+P_1, P_2, \ldots P_j\}$ and is hence not maximal in $o(j+1)$
(it would correspond to $Q=0$ in eq.~(\ref{5.9}) which is not allowed).

Case 6 can also be transformed into the MASA of eq.~(\ref{5.9}),
i.e. (\ref{5.15}) by a transformation of the form (\ref{5.21}) with
$G_0$ satisfying
\bea
G_0=\left( \begin{array}{cc} b & 1 \\ c & d  \end{array} \right),
\qquad
b+a_1 \not= 0, \qquad (k-1)c+d(a_2+\ldots +a_k)+md=0
\eea
Thus, all MASAs of $e(k,1)$ discussed above in points 4, 5 and 6 are
special cases of the free rowed MASA (\ref{5.9}) of $o(j+1,2)$ with
Kravchuk signature {\it (2 j-1 2)}. To determine the decomposition of
the space $M(j,1)$, consider a general transformation of the type
(\ref{5.21}). The entries depending on $\alpha$ in the first two rows
				       of $X$ transform as
\bea \label{5.28}
\left( \begin{array}{cc} a& b \\c& d \end{array} \right) 
\left( \begin{array}{c} \alpha  \\ \alpha Q \end{array} \right)= 
\left( \begin{array}{c}\alpha (a+bQ)  \\ \alpha (c+dQ) \end{array}
\right), \qquad
ad-bc \not = 0
\eea
We have
\bea \label{5.29}
a+bQ=diag(a+bq_1,a+bq_2, \ldots a+bq_{j-2})
\eea

To obtain a decomposition we must annul as many as possible of the
elements in the diagonal matrix (\ref{5.29}) by an appropriate choice
of $a$ and $b$. This number is equal to the highest multiplicity of an
eigenvalue of the matrix $Q$. Since we have $TrQ=0$, the multiplicity
is at most $j-3$. Let us order the eigenvalues in such a manner that
the last entry in $Q$ has the highest multiplicity equal to $r$. We
then choose $a$ and $b$ in eq.~(\ref{5.29}) so that the matrix in
(\ref{5.28}) has the form
\bea \label{5.30}
\left( \begin{array}{c} \alpha^{'} \\ \beta^{'} \end{array} \right) =
\left( \begin{array}{cccccc} \alpha_2 & \ldots & \alpha_s & 0 & \ldots &
0 \\
r_2 \alpha_2 & \ldots & r_s \alpha_s & \beta_1 & \ldots & \beta_r
 \end{array} \right), \qquad r+s=j
\eea
i.e. the MASAs
\bea
\{P_0-P_1, P_2+r_2(L_{02}-L_{12}), \ldots , P_s+r_s(L_{0s}-L_{1s}),
P_{s+1}, \ldots P_{s+r} \} \\
r_j \not= 0, \qquad 2 \leq j \leq s, \qquad \sum_{i=2}^{s}r_i=0 \nonumber \\
r_2=1\geq |r_3| \geq \ldots \geq |r_s| >0. \nonumber
\eea
Each integer $s$ and set of numbers $(r_2, \ldots r_s)$ corresponds to
an $O(p+1,2)$ conjugacy class of MASAs of $e(p,1)$.

Finally, let us sum up the above results as a theorem. 
\begin{theorem}
A representative list of maximal Abelian subalgebras of the
pseudoeuclidean Lie algebra $e(p,1)$ that are mutually inequivalent
under the action of the conformal group $O(p+1,2)$ coincides with a
list of MASAs of $o(p+1,2)$ of the form 
\bea \label{5.32}
MASA[e(p,1)] \sim l[o(2)] \oplus M_j, \qquad j=p-2l
\eea
where $M_j$ is a MASA of $o(j+1,2)$ contained in the subalgebra
$e(j,1)$. Specifically we have:
\begin{zoznamnum}
\item
$M_j \sim o(1,1) \oplus M_0$ where $M_0$ is a free rowed MANS of
$o(j,1)$ with Kravchuk signature {\it (1 j-1 1)} as in
eq.~(\ref{masaop1}).
The MASA of $e(p,1)$ is 
\bea
\{L_{12}, L_{34}, \ldots , L_{2l-1,2l}\} \oplus \{P_{2l+1},
\ldots , P_{p-1}\} \oplus \{L_{0p}\} \label{5.33}
\eea
\item
$M_j$ is a free rowed MANS of $o(j+1,2)$ with Kravchuk signature {\it (1 j+1 1)} as in
eq.~(\ref{5.7}). The MASA of $e(p,1)$ is
\bea \label{5.34}
\{L_{12}, L_{34} \ldots ,L_{2l-1,2l}\} \oplus \{P_0,P_{2l+1}, 
\ldots P_{p}\} 
\eea
\item
$M_j$ is a non-free rowed MANS of $o(j+1,2)$ with Kravchuk signature {\it (1~j+1~1)}
as in eq.~(\ref{5.8}). The MASA of $e(p,1)$ is 
\bea \label{5.35}
\{L_{12}, \ldots , L_{2l-1,2l}\} \oplus
\{L_{0,2l+1}-L_{p,2l+1}+\epsilon (P_0+P_p), P_0-P_p,\nonumber \\
 P_{2l+2}, \ldots
,P_{p-1}\}, \qquad \epsilon = \pm 1
\eea
\item
$M_j$ is a free-rowed MANS of $o(j+1,2))$ with Kravchuk signature {\it (2 j-1 2)}
as in eq.~(\ref{5.9}). The MASA of $e(p,1)$ is 
\bea \label{5.36} 
\{L_{12},\ldots,L_{2l-1,2l}\} \oplus
\{P_{2l+1}+q_{2l+1}(L_{0,2l+1}-L_{p,2l+1}), \ldots , \nonumber \\
P_{p-1}+q_{p-1}(L_{0,p-1}-L_{p,p-1}),P_0-P_p \}.
\eea
\end{zoznamnum}
Algebra (\ref{5.33}) is conformally equivalent to 
\bea \label {5.37}
\{L_{12}, \ldots ,L_{2l-1,2l} \} \oplus
\{P_0-P_p,(L_{0,2l+1}-L_{p,2l+1})+a_{2l+1}P_{2l+1}, \ldots , \nonumber \\
(L_{0s}-L_{ps})+a_sP_s,P_{s+1}, \ldots , P_{p-1}\} \\
r+s=j, \qquad \sum_{k=2l+1}^{s}a_k=0, \qquad a_{2l+1}=1 \geq
|a_{2l+2}| \geq \ldots \geq |a_s| >0
\eea
where $p-s-1$ is the highest multiplicity of any of the numbers
$q_{2l+1}, \ldots,q_p$.
\end{theorem}

Let us give some examples of the last  case in Theorem 5.1 for
$e(5,1)$.

\begin{zoznamrom}
\item $\{ P_{0}-P_{1}, L_{02}-L_{12},L_{03}-L_{13} \} \oplus L_{45}$ ($j=3$)\\
It can be represented as follows
\bea 
M=\left( \begin{array}{cccccccc} 0 & a & \ & \ & \ & \ & \ & \ \\
                                     -a & 0 & \ & \ & \ & \ & \ & \ \\
                                      \ & \ & 0 & 0 & 0 & 0 & d & 0 \\
                                      \ & \ & 0 & 0 & b & c & 0 & -d \\
                                      \ & \ & \ & \ & 0 & 0 & -b & \ \\
                                      \ & \ & \ & \ & 0 & 0 & -c & \ \\
                                      \ & \ & \ & \ & \ & \ & 0 & 0 \\
                                      \ & \ & \ & \ & \ & \ & 0 & 0 
               \end{array} \right), \label{5.38} \\
K=\left( \begin{array}{cccc} I_2 & \, & &  \\
                               & & & J_2 \\
                             \, & & I_2 & \, \\
                            & J_2 & \, & \, 
           \end{array} \right), \qquad
J_2=\left( \begin{array}{cc} 0 & 1 \\
                              1 & 0 
           \end{array} \right)  \nonumber 
\eea
which is equivalent under $O(6,2)$ to 
\bea M'=\left( \begin{array}{cccccccc} 0 & a & \ & \ & \ & \ & \ & \ \\
                                     -a & 0 & \ & \ & \ & \ & \ & \ \\
                                      \ & \ & 0 & 0 & b & c & -d & 0 \\
                                      \ & \ & 0 & 0 & 0 & 0 & 0 & d \\
                                      \ & \ & \ & \ & 0 & 0 & 0 & -b \\
                                      \ & \ & \ & \ & 0 & 0 & 0 & -c \\
                                      \ & \ & \ & \ & \ & \ & 0 & 0 \\
                                      \ & \ & \ & \ & \ & \ & 0 & 0 
               \end{array} \right),
\eea
$K$ is same as in (\ref{5.38}). This algebra is
$\{L_{45},P_0-P_1,P_2,P_3\}$ and is not maximal in $e(5,1)$ since we
can add $\{P_0+P_1\}$.

\item $\{P_{0}-P_{1},L_{02}-L_{12},L_{03}-L_{13}\} \oplus
\{P_{4},P_{5}\}$ ($j=5$) \\
It can be represented as
\bea M=\left( \begin{array}{cccccccc} 0 & 0 & a & b & 0 & 0 & e & 0 \\
                                      0 & 0 & 0 & 0 & c & d & 0 & -e \\
                                      \ & \ & 0 & 0 & 0 & 0 & 0 & -a \\
                                      \ & \ & 0 & 0 & 0 & 0 & 0 & -b \\
                                      \ & \ & \ & \ & 0 & 0 & -c & \ \\
                                      \ & \ & \ & \ & 0 & 0 & -d & \ \\
                                      \ & \ & \ & \ & \ & \ & 0 & 0 \\
                                      \ & \ & \ & \ & \ & \ & 0 & 0 
               \end{array} \right), \qquad
K=\left( \begin{array}{ccc} \, & \, &  J_2 \\
                             \, & I_4 & \, \\
                             J_2 & \, & \, 
           \end{array} \right).
\eea
This is equivalent under $O(6,2)$ to a 
\bea M'=\left( \begin{array}{cccccccc} 0 & 0 & a & b & c & d & e & 0 \\
                                      0 & 0 & -a & -b & c & d & 0 & -e \\
                                      \ & \ & 0 & 0 & 0 & 0 & a & -a \\
                                      \ & \ & 0 & 0 & 0 & 0 & b & -b \\
                                      \ & \ & \ & \ & 0 & 0 & -c & -c \\
                                      \ & \ & \ & \ & 0 & 0 & -d & -d \\
                                      \ & \ & \ & \ & \ & \ & 0 & 0 \\
                                      \ & \ & \ & \ & \ & \ & 0 & 0 
               \end{array} \right), 
\eea
$
M' \sim \{P_0-P_1,L_{02}-L_{12}-P_2, L_{03}-L_{13}-P_3,L_{04}-L_{14}
+P_4,L_{05}-L_{15}+P_5\}.$ \\
We see that here we have a free rowed MANS of $o(6,2)$ with Kravchuk signature
{\it ( 2 4 2)}. 
\item
$\{P_0-P_1,
L_{02}-L_{12}+P_2,L_{03}-L_{13}+aP_3,L_{04}-L_{14}-(1+a)P_4,L_{05}-L_{15}\}
\sim M$. This algebra is conformally equivalent to $M' \sim
\{P_0-P_1,P_2+L_{02}-L_{12},
P_3+a(L_{03}-L_{13}), P_4-(1+a)(L_{04}-L_{14})\}$ and will hence not
figure in the list given in Theorem 5.1 (i.e. $M'$ will figure, M will not).

\end{zoznamrom}



\section{Separation of variables in Laplace and wave operators}
\setcounter{equation}{0}
\subsection{MASAs and ignorable variables}

Let us consider an n-dimensional Riemannian, or pseudo-Riemannian space 
with metric
\bea
ds^2=g_{ik}(x)dx^idx^k \label{6.1}
\eea
and isometry group $G$. The Laplace-Beltrami equation on this space is
\bea \label{6.2}
\Delta_{LB}\Psi&=&E\Psi \nonumber \\
\Delta_{LB}&=&g^{-1/2}\sum_{i,j=1}^n {\partial \over \partial x^j}g^{1/2}g^{ij}
{\partial \over \partial x^i}, \qquad g=det(g_{ij})
\eea
and the Hamilton-Jacobi equation is
\bea \label{6.3}
g^{ij}{\partial S \over \partial x^i}{\partial S \over \partial x^j} =E.
\eea

We shall be interested in multiplicative separation of variables for eq.(\ref{6.2})
and additive separation for eq.(\ref{6.3}), i.e. in solutions of the form
\bea 
\Psi(x)=\prod_{i=1}^n \psi_i(x_i, c_1, \ldots , c_n), \label{6.4} \\
S(x)=\sum_{i=1}^n S_i(x_i, c_1, \ldots, c_n), \label{6.5}
\eea
respectively. Here $c_j$ are parameters, the separation constants and $\psi_i$ 
and $S_i$ obey ordinary differential equations.

A variable $x_j$ is {\it ignorable} \cite{eisen} if it does not figure in the 
metric tensor $g_{ik}$. Ignorable variables are directly related to elements of the Lie 
algebra 
L of the isometry group G \cite{milpatw}. Indeed, let ${X_1, \ldots X_l} \in L$ 
be a basis for an Abelian subalgebra of $L$. We can represent these elements by vector 
fields on $M$ expressed in terms of the coordinates $x$. Let us further assume 
that these vector fields are linearly independent at a generic point $x \in M$. 
We can then introduce (locally) coordinates on $M$

\bea
(x_1, \ldots, x_n) \longrightarrow ( \alpha_1, \ldots, \alpha_l, s_1, \ldots, s_k),
\qquad l+k=n \label{6.6}
\eea
that "straighten out" this algebra 
\bea \label{6.7}
X_i= {\partial \over \partial \alpha_i}, \qquad i=1, \ldots, l.
\eea
The variables $\alpha_i$ are the ignorable separable variables \cite{milpatw,eisen}.
Each MASA of the isometry algebra $L$ will provide a maximal set of ignorable 
variables, both for the Laplace-Beltrami and the Hamilton-Jacobi equation.

Specifically for the spaces $M(p,q)$ of this article, we generate the coordinates 
as follows. We use the realization (\ref{G}) of the group $E(p,q)$ but restrict $H$ 
to be a maximal Abelian subgroup of $E\pq$. We have $G=<\exp X>$, where $X$ is one 
of the MASAs we have constructed. We then write
\bea
\left( \begin{array}{c}
 x \\ 1 
\end{array} \right) =e^X \left( \begin{array}{c} s \\ 1  \end{array}\right),
\qquad s \in \mR^{p+q} \label{ex}
\eea
where $s$ represents a vector in a subspace of $M\pq$ parametrized by nonignorable 
variables $(s_1, \ldots ,s_k)$, and $X$ is a MASA of $e(p,q)$, parametrized by a 
set of ignorable variables.


\subsection{Ignorable variables in Euclidean space M(p)}

For Euclidean space the above considerations are entirely trivial. In cartesian 
coordinates we have 
\bea \label{6.8}
\Delta = {\partial^2 \over \partial x_1^2} + \ldots + 
{\partial^2 \over \partial x_n^2}.
\eea
In view of Theorem ~\ref{th3.3} we split the space $M(p)$ into a direct sum of 
one and two-dimensional spaces. In each $M(1)$ we have a Cartesian coordinate 
$x_i$, corresponding to the translation $P_i$. In each subspace $M(2)$ we 
have  polar 
coordinates, e.g. $M_{12}={\partial \over \partial \alpha_1}$ corresponds to
\bea 
x_1=s_1 \cos\alpha_1 \nonumber \\
x_2=s_1 \sin\alpha_1
\eea
with $\alpha_1$ ignorable.


\subsection{Ignorable variables in Minkowski space M(p,1).}

Here the situation is much more interesting. In Cartesian coordinates we have
\bea \label{6.10}
\Box_{p,1}\Psi=E\Psi \nonumber \\
\Delta_{LB} \equiv \Box_{p,1} ={\partial ^2 \over \partial x_1^2} +\ldots + 
{\partial ^2 \over \partial  x_p^2}-{\partial ^2 \over \partial x_0^2}. 
\eea
Consider the decomposition (\ref{4.29}) in Theorem ~\ref{th4.3}. In each 
indecomposable subspace we introduce a separable system of coordinates with 
a maximal number of ignorable variables. Each space $M(1,0)$ corresponds to 
a Cartesian coordinate, $M(2,0)$ to polar coordinate as in eq. (\ref{6.8}). 
Now let us consider the coordinates corresponding to $M(k,1)$. 
\bea
\begin{array}{lcl}
M(0,1) & :&  x_0  \nonumber \\
M(1,1)  &:&  x_0=\rho \cosh \alpha, \, \, \, \, \, \, x_1=s\sinh\alpha   \\
        & &   x_0=\rho \sinh \alpha, \, \, \, \, \, \, x_2=s\cosh\alpha  \\
 & & ({\rm for} \, \, x_0^2-x_1^2= \pm s^2, {\rm respectively}) \nonumber \\
\end{array} \nonumber
\eea

\noindent M(2,1)  : The algebra (\ref{4.27}) with $\kappa=1$ provides two 
ignorable variables, 
$z$ and $a$ and we have \\
\bea \label{6spec}
\begin{array}{ccl}
x_0+x_1&=& r\sqrt{2}+2a  \\
x_0-x_1&=&ra^2\sqrt{2}+{2 \over 3}a^3 -z \sqrt{2}\\
x_2 & =& -a^2-ar\sqrt{2}. 
\end{array} \label{6.11}
\eea
The coordinates (\ref{6.11}) were obtained using eq.(\ref{ex}) with
\bea
G=e^X, \qquad X=\left( \begin{array}{cccc} 0 & a \sqrt{2} & 0 & z\sqrt{2} \\
                                           0 & 0 & -a \sqrt{2} & 0 \\
                                           0 & 0 & 0 & a \sqrt{2} \\
                                           0 & 0 & 0 & 0 
                        \end{array} \right), \qquad
s=\left( \begin{array}{c} 0 \\ 0 \\ r \end{array} \right).
\eea
We then have
\bea
P_0-P_1=- {\partial \over \partial z}, \qquad L_{02}-L_{12}+P_0+P_1=
{\partial \over \partial a}
\eea
and the operator in this $M(2,1)$ subspace of $M(p,1)$ is:
\bea
\Box_{2,1}=\sqrt{2}{\partial^2 \over \partial r \partial z}+{1 \over 2}{1 \over r^2}
{\partial^2 \over \partial a^2}+{1 \over r^2}{\partial \over \partial r^2}-
{\sqrt{2} \over r^2}{\partial^2 \over \partial r \partial a} \nonumber \\+
{1 \over \sqrt{2}}{1 \over r}{\partial \over \partial z}-{1 \over r^3}{\partial \over \partial r}+
{1 \over \sqrt{2}}{1 \over r^3}{\partial \over \partial a}
\eea

The separated solutions of the wave equation (\ref{6.10}) have  the  form:
\bea
\Psi=R_{Eml}(r)e^{mz}e^{la}
\eea
The equation for $R_{Eml}(r) \equiv R$ has the form 
\bea
R^{''}+\tilde p(r)R^{'}+\tilde q(r)R=0
\eea
Using the transformation
\bea
R(r)&=&f(r)W(\rho) \nonumber \\
f(r)&=&r^{{2-\lambda-\lambda^{'}} \over {2}} \exp \left( -{{mr^3} \over 3} 
+{{lr} \over \sqrt{2}} \right), \qquad \rho=r^{-2}
\eea
we obtain the equation
\bea
W^{''}+ p(\rho)W^{'}+ q(\rho)W=0  \label{Wspec}
\eea
where $ p(\rho)$ and $ q(\rho)$ are
\bea
 p(\rho)={{1-\lambda -\lambda^{'}} \over {r^{-2}}}, \qquad q(\rho)=-k^2
 +2\alpha r^2 +\lambda \lambda^{'} r^4  
\eea
\bea
\lambda^{'}={{(A-1) \pm \sqrt{(a-1)^2+4m^2}} \over {2}}, 
\qquad 1- \lambda - \lambda^{'} =A, \nonumber \\
 \qquad A=3 \, \, {\rm or} \, {1 \over 2}, \qquad 2\alpha= lm\sqrt{2}-E.
\eea
The solution of (\ref{Wspec}) is a confluent hypergeometric series \cite{morse}.

Let us consider the space $M(k,1)$ with $k \geq 2$  and the splitting MASA (\ref{4.28})
with $a_2=a_3= \ldots =a_k=0$. The corresponding matrix realization is given by 
eq. (\ref{masa})
with $M_0$ and $\gamma$ as in eq. (\ref{p1spl:k01}) and all the $M_i$ and $x$ absent. 
Applying eq.
(\ref{ex}) with
\bea
X=\left( \begin{array} {cccc} 0 & \alpha & 0 &z \\
                              0 & 0 & -\alpha^T & 0 \\
                               0&0&0&0 \\
                              0&0&0&0 
          \end{array} \right), \qquad s=\left( \begin{array} {c} 0 \\ \vdots \\ 0 \\ r 
                                                \end{array} \right), \qquad r \in \mR
\eea
we obtain the coordinates
\bea
\begin{array}{ccl} \label{6.13}
{x_k+x_0}  & = & r \sqrt{2}  \\
{x_k-x_0}  & = & -r \alpha \alpha^T {1 \over \sqrt{2}} + z \sqrt{2} \\
x_1 & = & -r \alpha_1  \\
& \vdots & \\
x_{k-1} & = & -r \alpha_{k-1} 
\end{array} 
\eea

The wave operator in these coordinates is
\bea \label{6.14}
\Box_{k,1}=2 {\partial^2 \over {\partial z \partial r}}+ {k-1 \over r} 
{\partial \over \partial z} + {1 \over r^2} \sum_{i=1}^{k-1} 
{\partial^2 \over {\partial \alpha_i^2}}.
\eea 

The variables $z$ and $\alpha_i$ are ignorable (only $r$ figures in eq.(\ref{6.14})) 
and indeed we have
\bea \label{6.15}
P_0-P_k=- \sqrt{2}{\partial \over \partial z}, \qquad L_{0i}-L_{ki}=\sqrt{2}
{\partial \over \partial \alpha_i}. 
\eea

The solution of the wave equation then separates 
\bea
\psi=R(r)e^{mz}\prod_{i=1}^{k-1}e^{b_i\alpha_i}
\eea
with $R(r)$ as follows
\bea
R(r)=r^{-k \over 2} \exp({1 \over r}{\sum_{i=1}^{k-1}b_i^2 \over 2m})
\exp({Er \over 2m}).
\eea

We have shown in Section 5.3 that this MASA is conformaly equivalent to a subalgebra 
of the algebra of translations, namely to $( P_0-P_k,P_1, \ldots , P_{k-1})$. 
A consequence of this is that we can relate these coordinates to a set of cartesian ones. 
Indeed, we can rewrite eq. (\ref{6.14}) as
\bea \label{6.16}
\Box_{k,1}=(y_0+y_k)^{k-1 \over 2}(y_0+y_k)^2 \left[ {{\partial^2} \over 
{\partial y_0^2}}-
{{\partial^2} \over {\partial y_1^2}}- \ldots -{{\partial^2} \over 
{\partial y_{k}^2}} \right] (y_0+y_k)^{-{k-1 \over 2}}
\eea
with
\bea 
\begin{array}{ccl} \label{6.17}
{x_1+x_0} & =& - { 1 \over {y_0+y_k}} \sqrt{2} \\
{x_1-x_0} & =& -{1 \over \sqrt{2}}{1 \over {y_0+y_k}} (y_0^2-y_1^2-\ldots -y_{k}^2) \\
x_j &=& {y_j \over {y_0+y_k}}, \qquad j=1, \ldots k-1.
\end{array} 
\eea
We note however that the wave equation separates in coordinates $(r,z,\alpha_i)$ 
but not
in $(y_0,y_1, \ldots ,y_k)$.

Now consider the space $M(k,1)$ for $k \geq 3$ and the nonsplitting MASA (\ref{4.28}) 
with $a_i \not= 0$. The coordinates we obtain are
\bea \label{6.18}
\begin{array}{ccl}
x_k+x_0  & = & r \sqrt{2} \\
x_k-x_0 & = & {1 \over \sqrt{2}}(2z-r \alpha \alpha^T +\alpha A \alpha^T) \\
x_1 & = & (q_1-r) \alpha_1 \\
& \vdots & \\
x_{k-1} & = & (q_{k-1}-r) \alpha_{k-1}. 
\end{array} 
\eea

The wave operator is
\bea \label{6.19}
\Box_{k,1}=2 {\partial^2 \over {\partial z \partial r}} - 
\left( \sum_{i=1}^{k-1} {1 \over {(q_i-r)}}\right) {\partial \over \partial z} 
+  \sum_{i=1}^{k-1} {1 \over {(q_i-r)^2}}
{\left( \partial^2 \over {\partial \alpha_i^2}\right)}.
\eea 
We see that $\alpha_k, z$ are ignorable variables.
The solution of the wave equation then separates and we have
\bea
\Psi=R(r)e^{mz}\prod_{i=1}^{k-1}e^{a_i\alpha_i}
\eea
with $R(r)$ equal to
\bea
R(r)=\prod_{i=2}^{k} (q_i-r)^{-{1 \over 2}} \exp {\left( -{1 \over 2m} \sum_{i=2}^{k}
{b_i^2 \over {q_i-r}} \right)}
\exp{ \left( {Er \over 2m} \right)}. 
\eea

We mention that the three new coordinates systems, (\ref{6spec}), (\ref{6.13}) and 
(\ref{6.18})
are all nonorthogonal, hence the cross terms (mixed derivatives) in the corresponding 
forms 
of the wave operator.


\section {Conclusions}
\setcounter{equation}{0}

The classification of MASAs of $e(p,0)$ and $e(p,1)$ performed in this article is 
complete, entirely explicit and the results are reasonably simply. Indeed, they are 
summed up in Theorems 3.1 and 3.2 and 3.3 for $e(p,0)$ and Theorems 4.1, 4.2, 4.3 
and 5.1 for $e(p,1)$.

In Section 6 we have presented a first application of this classification. Namely, 
we have constructed the coordinate systems (\ref{6.11}), (\ref{6.13}) and (\ref{6.18}) 
which allow the separation of variables in the wave equation and have the maximal 
number of ignorable variables. In turn, these coordinate systems have further applications.

Thus, instead of the wave equation itself, let us consider a more general equation, namely
\bea \,
[\Box+V(x)]\Psi=E\Psi \label{7.1}.
\eea
First of all, it is possible to choose the potential V(x) to be such that 
eq.(\ref{7.1})
allows the separation of variables in one of the above coordinate systems. The 
obtained equation will be integrable in that there will exist a complete set of 
$p$ second order operators commuting with $H=\Box+V$ and amongst each other. They 
will be of the form $X_i^2+f_i(x_i)$ where $\{X_i\}$ is corresponding MASA and 
$f_i(x_i)$ is a function of the corresponding ignorable variable. The actual form 
of $f$ depends on the separable potential $V(x)$ \cite{uhl,mak}.

The coordinates (\ref{6.18}) have been used to construct equations of the type 
(\ref{7.1}) that obey the Huygens principle \cite{Ber}. Crum-Darboux transformation 
\cite{Crum} \cite{Dar} \cite{Matv} can
be used to generate specific potentials $V(x)$ (depending on one ignorable variable 
in a given separable coordinate system) that have specific integrability properties. 
In particular this proviodes a method for constructing overcomplete commutative rings 
of partial differential operators and "algebraically integrable" dynamical systems 
\cite{Krich}, \cite{Chal}, \cite{Ves}.

The reason we bring this up here is that traditionally the Crum-Darboux transformations 
have been performed in cartesian, or polar coordinates. The fact that they can be 
applied to other types of coordinates, associated to other types of MASAs, opens new 
possibilities.

Work is in progress on the classification of MASAs of $e\pq$ for $p \geq q \geq 2$ \cite{ZW}.

\bigskip
\medskip

{\Large\textbf{Acknowledgements}} \\

\medskip

We thank Yu. Berest and I. Lutsenko for very helpful discussions. 
The research of P.W. is partially supported by research grants from NSERC of Canada 
and FCAR du Qu{\'e}bec.
Z.T. was partially suported by Bourse de la F.E.S., Universit{\'e} de Montr{\'e}al.

\bibliography{CRM2473}

\begin{thebibliography}{10}

\bibitem{olver}
P.J. Olver.
\newblock {\em Applications of {L}ie groups to differential equations}.
\newblock Springer-Verlag, New York, 1993.

\bibitem{kluwer}
P.~Winternitz.
\newblock Lie groups and solutions of partial differential equations.
\newblock In A.~Ibort and M.A. Rodriguez, editors, {\em Integrable systems,
  quantum groups and quantum field theories}. Kluwer, Dordrecht, 1993.

\bibitem{frisw}
P.~Winternitz and I.~Fri{\v s}.
\newblock Invariant expansions of relativistic amplitudes and the subgroups of
  the proper {L}orentz group.
\newblock {\em Yad. Fiz}, pages 889--901, 1965.
\newblock [Sov. J. Nucl. Phys., 1:636-643, 1965].

\bibitem{wsmor}
P.~Winternitz, I.~Luka{\v c}, and Y.A. Smorodinskii.
\newblock Quantum numbers in the little group of the {P}oincar{\'e} group.
\newblock {\em Yad. Phys.}, 7:192--201, 1968.
\newblock [Sov. J. Nucl. Phys., 7:139-145, 1968].

\bibitem{miller}
W.~Miller~Jr.
\newblock {\em Symmetry and Separation of Variables}.
\newblock Addison-Wesley, Reading,Mass., 1977.

\bibitem{kalnins}
E.G. Kalnins.
\newblock {\em Separation of Variables for Riemannian Spaces of Constant
  Curvature}.
\newblock Pitman, New York, 1986.

\bibitem{milpatw}
W.~Miller~Jr., J.~Patera, and P.~Winternitz.
\newblock Subgroups of {L}ie groups and separation of variables.
\newblock {\em J.Math.Phys.}, 22:251--260, 1981.

\bibitem{eisen}
L.P. Einsenhart.
\newblock Separable systems of {S}t{\"a}ckel.
\newblock {\em Ann. Math.}, 35:284--305, 1934.

\bibitem{sp_C}
J.~Patera, P.~Winternitz, and H.~Zassenhaus.
\newblock Maximal {A}belian subalgebras of real and complex symplectic {L}ie
  algebras.
\newblock {\em J.Math.Phys.}, 24:1973--1985, 1983.

\bibitem{su_pq}
M.A. Olmo, M.A. Rodriguez, and P.~Winternitz.
\newblock Maximal {A}belian subalgebras of pseudounitary {L}ie algebras.
\newblock {\em Linear Algebra Appl.}, 135:79--151, 1990.

\bibitem{verc}
V.~Hussin, P.~Winternitz, and H~Zassenhaus.
\newblock Maximal {A}belian subalgebras of complex orthogonal {L}ie algebras.
\newblock {\em Linear Algebra Appl.}, 141:183--220, 1990.

\bibitem{verop}
V.~Hussin, P.~Winternitz, and H.~Zassenhaus.
\newblock Maximal {A}belian subalgebras of pseudoorthogonal {Lie} algebras.
\newblock {\em Linear Algebra Appl.}, 173:125--163, 1992.

\bibitem{jacobson}
N.~Jacobson.
\newblock {\em Lie algebras}.
\newblock Dover, New York, 1079.

\bibitem{kostant}
B.~Kostant.
\newblock On the conjugacy of real {C}artan subalgebras {I}.
\newblock {\em Proc. Nat. Academy Sci. USA}, 41:967--970, 1955.

\bibitem{sugiura}
M.~Sugiura.
\newblock Conjugate classes of {C}artan subalgebras in real semi-simple
  algebras.
\newblock {\em J. Math. Soc. Japan}, 11:374--434, 1959.

\bibitem{suprtysh}
D.A. Suprunenko and R.I. Tyshkevich.
\newblock {\em Commutative matrices}.
\newblock Academic Press, New York, 1968.

\bibitem{maltsev}
A.I. Maltsev.
\newblock Commutative subalgebras of semi-simple {L}ie algebras.
\newblock {\em Izv. Akad. Nauk SSR Ser. Mat}, 9:291, 1945.
\newblock Amer. Math. Soc. Transl. Ser. 1 9:214 (1962).

\bibitem{laffey}
T.J. Laffey.
\newblock The minimal dimension of maximal commutative subalgebras of full
  matrix algebras.
\newblock {\em Linear Algebra Appl.}, 71:199--212, 1985.

\bibitem{kal}
E.G Kalnins and P.~Winternitz.
\newblock Maximal {A}belian subalgebras of complex euclidean {L}ie algebras.
\newblock {\em Can. J. Phys.}, 72:389--404, 1994.

\bibitem{morse}
P.M. Morse and H.~Feshbach.
\newblock {\em Methods of Theoretical Physics}.
\newblock McGraw-Hill, New York, 1953.

\bibitem{uhl}
P.~Winternitz, Ya.A. Smorodinsky, M.~Uhl{\'\i}{\v r}, and I.~Fri{\v s}.
\newblock Symmetry groups in classical and quantum mechanics.
\newblock {\em Yad. Fiz.}, 4:625--635, 1966.

\bibitem{mak}
A.~Makarov, Ya. Smorodinsky, Kh. Valiev, and P.~Winternitz.
\newblock A systematic search for nonrelativistic systems with dynamical
  symmetries.
\newblock {\em Nuovo Cim.}, A 52:1061--1084, 1967.

\bibitem{Ber}
Yu.Yu. Berest and P.~Winternitz.
\newblock Huygens' principle and separation of variables.
\newblock 1996.
\newblock preprint CRM-2379, to be published.

\bibitem{Crum}
M.~Crum.
\newblock Associated {S}turm-{L}iouville systems.
\newblock {\em Quart. J. Math}, 6(2):121--127, 1955.

\bibitem{Dar}
G.~Darboux.
\newblock Sur la representation sph{\' e}rique des surfaces.
\newblock {\em Compt. Rendus}, 94:1343--1345, 1882.

\bibitem{Matv}
V.B. Matveev and M.A. Salle.
\newblock {\em Darboux Transformations and Solitons}.
\newblock Springer-Verlag, Berlin, 1991.

\bibitem{Krich}
I.M. Krichever.
\newblock Methods of algebraic geometry in the theory of nonlinear equations.
\newblock {\em Russian Math. Surveys}, 32(6):198, 1977.

\bibitem{Chal}
O.A. Chalykh and A.P. Veselov.
\newblock Commutative rings of partial differential operators and {L}ie
  algebras.
\newblock {\em Comm. Math. Phys}, 126:597, 1990.

\bibitem{Ves}
A.P. Veselov.
\newblock Huygens' principle and algebraic {S}chr{\"o}dinger operators.
\newblock In {\em Topics in Topology and Mathematical Physics}, volume 102,
  pages 199--206. Amer. Math. Soc. Transl., Ser. 2, 1995.

\bibitem{ZW}
Z.~Thomova and P.~Winternitz.
\newblock Maximal {A}belian subalgebras of pseudoeuclidean {L}ie algebras.
\newblock To be published.

\end{thebibliography}
\end {document}